\documentclass[a4paper,useAMS,usenatbib,usegraphicx]{mn2e}
\usepackage{multirow}
\usepackage{aas_macros}
\usepackage{color}
\usepackage[total={17.8cm,24.0cm},centering]{geometry}
\usepackage{times}
\usepackage{amssymb}

\newcommand{\sauron}{{\texttt {SAURON}}\ }

\title[Distinct stellar populations in inner bars]{%
Distinct stellar populations in the inner bars of double-barred galaxies}
\author[de Lorenzo-C\'aceres et al.]{%
A. de Lorenzo-C\'aceres$^{1,2}$\thanks{adlcr@iac.es}, 
J. Falc\'on-Barroso$^{1,2}$\thanks{jfalcon@iac.es}, and 
A. Vazdekis$^{1,2}$\thanks{vazdekis@iac.es}\\
$^{1}$Depto. Astrof\'isica, Universidad de La Laguna (ULL), E-38206 La Laguna, Tenerife, Spain\\
$^{2}$Instituto de Astrof\'isica de Canarias (IAC), E-38205 La Laguna, Tenerife, Spain}

\begin{document}

\date{Accepted xxx. Received xxx; in original form xxx}
\pagerange{\pageref{firstpage}--\pageref{lastpage}} \pubyear{2002}

\maketitle

\label{firstpage}

\begin{abstract}
Numerical simulations of double-barred galaxies predict the build-up of
different structural components (e.g., bulges, inner discs) in the central
regions of disc galaxies. In those simulations, inner bars have a prominent
role in the internal secular evolution of their host galaxies. The development
of bulges and inner discs is, however, poorly understood observationally due to the
small number of studies focusing on the stellar populations of these systems.
In order to provide constraints on the relevant processes inducing the creation
of these components in the presence of inner bars, we have carried out a
detailed kinematical and stellar population analysis in a sample of four
double-barred galaxies, ranging from SB0 to SBb,
observed with integral-field spectroscopy. We find that
the inner bars present distinct stellar population properties, being younger and
more metal-rich than the surrounding bulges and outer bars. 
While we detect signatures of gas inflow through the inner bars,
we find no evidence of associated star-forming regions or newly-formed structures
around them. This result suggests that, regardless of their formation scenario,
\textit{at present} these inner bars are
playing a moderate or even a minor role in the morphological evolution of this
sample of double-barred galaxies.
\end{abstract}


\section{Introduction}

Bars have been traditionally considered key drivers of the internal secular
evolution of disc galaxies. In fact, bars are non-axisymmetric structures that
can redistribute the angular momentum of a galaxy, thus favouring the transport
of gas through them to the inner regions where it may trigger star
formation \citep[e.g.,][]{MunozTunonetal2004,Shethetal2005}. This internal, slow
process competes with fast external drivers, such as mergers or interactions, as
the main scenarios responsible for the formation of new spheroidal components
and bulges at the galaxy centres \citep[see][ for a
review]{KormendyandKennicutt2004}. Moreover, bar-driven structural evolution of
disc galaxies can produce significant kinematical and morphological changes,
such as the increase of the bulge-to-disc ratio, which in turn could lead to a
parallel evolution along the Hubble sequence
\citep{PfennigerandNorman90,FriedliandMartinet93}.

The bar-driven evolution scenario is backed by the large fraction of barred
galaxies found in the nearby Universe
\citep[e.g.,][]{Eskridgeetal2000,MarinovaandJogee2007,Aguerrietal2009} to
$z\sim1$ (e.g., \citealt{Elmegreenetal2004,Jogeeetal2004}, but see also
\citealt{Shethetal2008,Cameronetal2010}). Among local early-type 
barred spirals, $\sim$30\% have been found to host an inner, secondary bar embedded in the main,
larger bar \citep[e.g.,][]{ErwinandSparke2002,Laineetal2002,Erwin2004}, and a
two-bar system has been observed even at $z\sim0.15$ by \citet{Liskeretal2006}.
Therefore, double-barred galaxies are also rather common systems and, in the
same way as a single bar, they may contribute to the gas inflow towards the
central regions. Moreover, the material driven inwards by 
a nested-bar system may
reach the innermost parts and feed the active galactic nuclei \citep[][ but see
also \citealt{Hoetal97,Marquezetal2000}]{Shlosmanetal89,Shlosmanetal90}. It is
hence clear that inner bars may be playing an important role in the shaping
and evolution of spiral galaxies.

The formation of inner bars and their efficiency in driving the gas inflow have been
extensively studied by numerical simulations. These works usually require the
presence of a dissipative component to form a stable double-barred system, so
the main bar is formed first, 
and the gas flows along it and is finally trapped by
the $x_2$ orbits. The $x_2$ family is composed by elliptical orbits
which are elongated in the direction perpendicular
to the main bar major axis \citep{Contopoulosetal80}. 
The gas trapped by the $x_2$ orbits then forms the inner bar
\citep{FriedliandMartinet93,Helleretal2001,Rautiainenetal2002,
ShlosmanandHeller2002, EnglmaierandShlosman2004}. Only
\citet{DebattistaandShen2007} are able to create a purely stellar inner bar from
collisionless $N$-body simulations, but with the peculiar initial requirement
of having  a rapidly rotating component at the centre. Except for this
particular case, most numerical simulations produce gas-rich bars\linebreak capable of
transporting material to the central regions \citep[e.g.,][]{Helleretal2007},
thus contributing significantly to the secular evolution of their host
galaxies \citep{PfennigerandNorman90}.\looseness-1

In an effort to constrain the role of bar-related secular processes in
shaping the central regions of galaxies, several authors have compared the
properties of bulges in barred and non-barred galaxies.
\citet{MoorthyandHoltzman2006} analysed a sample of face-on spirals 
and obtained a
trend toward younger ages for the bulges of barred galaxies; on the contrary,
\citet{Jablonkaetal2007} found no differences between the stellar population
properties of edge-on barred and non-barred galaxies. \citet{Perezetal2011}
studied the bulge properties of a sample of early-type galaxies hosting bars,
and found significant differences with respect to their unbarred counterparts, the
former being more metal-rich and slightly more $\alpha$-enhanced. The
metallicity of the bulge was very similar to that of the bar, and
\citet{Perezetal2011} concluded that both structures form simultaneously but with
different episodes of star formation. Finally, \citet{CoelhoandGadotti2011}
compared the age and metallicity values for a sample of bulges hosted by barred
and non-barred galaxies; their results suggested that bars do affect the mean
stellar age of bulges, in such a way that bars actually triggered star formation
at the centres of their galaxies.

None of the previous studies includes specific data for double-barred galaxies.
In fact, there is a notable lack of kinematical and stellar population analyses
of these kinds of objects in the literature, mainly due to the strict
observational conditions needed to take spectra of these structurally complex
objects, which are composed of two bars randomly oriented
\citep[e.g.,][]{FriedliandMartinet93}, very different in size
\citep{ErwinandSparke2002}, and independently rotating
\citep[e.g.,][]{ShlosmanandHeller2002,Corsinietal2003}. \citet{Perezetal2007},
\citet{Perezetal2009}, and \citet{SanchezBlazquezetal2011} studied the stellar
populations of a sample of barred galaxies including some double-bars; they found
positive, null, and negative metallicity gradients along the bars, independently
of their age profiles, and also concluded that bulges of barred galaxies tend to
be more metal-rich than those belonging to unbarred galaxies of similar velocity
dispersions. However, these authors focussed on the main bars and did
not pay special attention to the behaviour of inner bars. Only
\citet{deLorenzoCaceresetal2012} studied in detail the stellar population
properties of the bulge, inner bar, and outer bar of a double-barred galaxy:
NGC\,357. They found that the three structures are nearly coeval, with an age
around 8\,Gyr; the bulge and inner bar also show the same metallicity and
$\alpha$-enhancement values, but the outer bar is clearly less metal-rich and
more [Mg/Fe] overabundant than the inner regions. 
The interpretation of these results was that the inner bar and bulge
were effected by the redistribution of existing stars.

The analysis performed by \citet{deLorenzoCaceresetal2012} relied on long-slit
spectroscopy along both bars, so their results could not address the regions
outside the bars
that might be quite important given the structural complexity of double-barred
galaxies. Integral-field spectroscopy seems to be the most suitable option for
studying these kinds of objects, as shown by \citet{Moiseev2001},
\citet{Moiseevetal2004}, and \citet{deLorenzoCaceresetal2008}. These three works
are focused on the kinematics of double-barred galaxies, previously studied
through long-slit spectroscopy by \citet{Emsellemetal2001}, and they highlight
the importance of including the spatial information outside the major
axes of the bars in the analysis. In
particular, \citet{deLorenzoCaceresetal2008} presented the discovery of the
$\sigma$-hollows, two local decreases of the velocity dispersion
values exactly at the edges of the inner bars. With the aid of numerical
simulations, the authors infer that these hollows are due to the contrast
between the different velocity dispersion values of the bulge and the inner bar,
and they represent the only known kinematical signature of the presence of a
stellar inner bar. It is important to notice that although the $\sigma$-hollows
have also been observed in long-slit spectra \citep{deLorenzoCaceresetal2012},
their discovery is closely linked to the integral-field nature of the work
performed by \citet{deLorenzoCaceresetal2008}.

We present here a detailed kinematical and stellar population analysis of the
sample of double-barred galaxies, already introduced in
\citet{deLorenzoCaceresetal2008}, studied through integral-field spectroscopy.
For the first time, we analyse the spatial distribution of the age, metallicity,
abundance ratio, and other relevant properties of double-barred galaxies in
order to constrain their formation scenarios and role in secular evolution.
The paper is organised as follows: in Section \ref{sec:sample} we describe the
sample selection and observations, whereas the data reduction is summarized
in Section \ref{sec:reduction}. In Section \ref{sec:phot} we perform a
photometric analysis to check the parameters of the inner and outer bars.
Extensive descriptions of the kinematical and stellar population analyses, and
the direct results obtained from them, can be found in Sections \ref{sec:kin}
and \ref{sec:sp}, respectively. The implications of these results for the
formation scenarios of double-barred galaxies and the role of inner bars in the
galaxy evolution are discussed in Section \ref{sec:discussion}. Finally, a
summary of the work and the main conclusions are compiled in Section
\ref{sec:conclusions}.

\section{Sample selection, observations and instrumental setup}
\label{sec:sample}

\begin{table*}
\begin{center}
\caption{Main properties of the double-barred sample.}
\label{tab:sample}
\begin{tabular}{ccccccccccc}
\hline
Name      & RC3 Type      & \emph{D} & \emph{i}    & \multicolumn{3}{c}{Position angle}                 & \multicolumn{2}{c}{Semi-major axis} & \multicolumn{2}{c}{$\epsilon_{max}$}\\  
(1)       & (2)           & (3)      & (4)         & \multicolumn{3}{c}{(5)}                            & \multicolumn{2}{c}{(6)}             & \multicolumn{2}{c}{(7)} \\ 
\hline
          &               &          &             & Disc         & Inner bar     & Outer bar           & Inner bar  & Outer bar              & Inner bar & Outer bar    \\
\hline        
NGC\,2859 & (R)SB(r)0$^+$ & 25.4 Mpc & 25$^{\circ}$ & 90$^{\circ}$  & 62$^{\circ}$   & 162$^{\circ}$        & 4.1 arcsec & 34  arcsec             & 0.31      & 0.40  \\ 
NGC\,3941 & SB(s)0$^0$    & 18.9 Mpc & 51$^{\circ}$ & 10$^{\circ}$  & 30$^{\circ}$   & 166$^{\circ}$        & 3.2 arcsec & 21  arcsec             & 0.21      & 0.47  \\
NGC\,4725 & SAB(r)ab      & 12.4 Mpc & 42$^{\circ}$ & 40$^{\circ}$  & 141$^{\circ}$  & 50$^{\circ}$         & 5.6 arcsec & 118 arcsec             & 0.20      & 0.67  \\
NGC\,5850 & SB(r)b        & 28.5 Mpc & 30$^{\circ}$ & 163$^{\circ}$ & 50$^{\circ}$   & 116$^{\circ}$        & 5.9 arcsec & 63  arcsec             & 0.30      & 0.68  \\
\hline
\end{tabular}
\end{center}
\begin{minipage}{17cm}
(1) Galaxy name; (2) Morphological type from \citet{rc3}; (3) distance to the
galaxy, from \citet{Tully88}. Throughout the whole paper we assume
H$_0$=75\,km\,s$^{-1}$\,Mpc$^{-1}$; (4) galaxy  inclination; (5) position angle of the
major axis (disc) and the two bars; (6) bar lengths, estimated as the semi-major
axes of maximum isophotal ellipticity; (7) maximum isophotal ellipticity of the
bars ($\epsilon=1-b/a$). Columns (4), (5), (6), and (7) are taken from
\citet{Erwin2004} \citep[except for the PA of the inner bar of NGC\,3941, which
is taken from][]{ErwinandSparke2003}.
{\sc Note:} there is an erratum in \citet{Erwin2004} so that
the position angle of the inner bar of NGC\,3941 is given as 85$^{\circ}$, which
is not correct. \citet{ErwinandSparke2003} provide a value between 20$^{\circ}$
and 35$^{\circ}$; taking into account the alignment of the $\sigma$-hollows (see
Section \ref{sec:hollows}), we finally assume a position angle of 30$^{\circ}$
for NGC\,3941, in agreement with the measurements of
\citet{ErwinandSparke2003}.
\end{minipage}
\end{table*}

The sample is composed of four double-barred galaxies: NGC\,2859, NGC\,3941,
NGC\,4725, and NGC\,5850. These galaxies were selected from the catalogue of
\citet{Erwin2004}, paying special attention to the length and size of the inner
bars. The main requirement for building the sample was that the inner bars were
big enough to be well mapped within the \sauron\ field-of-view (hereafter FoV),
but they were also small enough to allow the mapping of the transition regions
between the two bars. Note that the main bar is usually larger than the total
FoV. 

The inclinations of the four galaxies range between 25$^{\circ}$ and
50$^{\circ}$; these intermediate values allow us to recover the line-of-sight
kinematics, with only a moderate mixing of the components due to projection.
Since the main goal is to perform the stellar population analysis, only
early-type galaxies were considered in order to avoid additional handicaps, such
as complex spiral structures. In particular, the galaxies were selected to
have almost no dust, although it is worth noting here that our stellar population analysis 
is based on the measurement of the line indices which are little affected by dust
\citep[see][]{MacArthur2005}.
Finally, NGC\,3941 and 
NGC\,4725 are classified as Seyfert 2 galaxies \citep{Veron-Cettyetal2006}, 
whereas NGC\,2859 presents no signs
of nuclear activity and NGC\,5850 is considered a LINER 
\citep[Low Ionisation Nuclear Emission Region, see
Section  \ref{sec:gas5850} for further details on the LINER classification
of this galaxy;][]{Bremeretal2012}. The bar lengths,
ellipticities, position angles (hereafter PA), and other relevant properties for
the sample galaxies are summarized in Table~\ref{tab:sample}. 

The four double-barred galaxies have neutral atomic gas, as shown in the HI
studies by \citet{Serraetal2012} for NGC\,2859 and NGC\,3941, 
\citet{Weversetal84} for NGC\,4725, and \citet{Higdonetal98} for NGC\,5850.
The HI is mainly located in external ring structures, whereas there is no
neutral atomic gas in the central regions we are interested in.
Regarding the neutral molecular gas, CO observations have been taken for 
NGC\,2859 \citep[][]{BiegingandBiermann77,WardleandKnapp86,PetitpasandWilson2004}, 
NGC\,3941 \citep[][]{Bontempietal2012}, and NGC\,4725 \citep[][]{Shethetal2005}.
If detected at all, the molecular gas fractions were very low.
Particularly interesting is the case of NGC\,5850, for which a very large amount of
molecular gas is found in the central region, possibly driven there through 
the main bar \citep[][]{PetitpasandWilson2004}.
Moreover, high resolution interferometric maps of NGC\,5850 obtained by 
\citet[][]{Leonetal2000} reveal that most of this central molecular gas is concentrated
in a single peak, which is slightly off-centred towards the North.

Our new observations were carried out with the \sauron\ integral-field spectrograph,
attached to the William Herschel Telescope (WHT) at the Observatorio del Roque
de los Muchachos in La Palma, Spain. Data for the four galaxies were taken
during four nights in 2007 April 14 and 16--18. Detailed information on the
observations is found in Table~\ref{tab:observations}. \sauron\ uses an
array of lenslets to section the FoV into 1431 
\emph{spaxels} (spatial elements). The spectrograph was
operated in the Low-Resolution (LR) mode, which provides a
$33\times41$\,arcsec$^2$ FoV with a spatial scale of 0.94\,arcsec~spaxel$^{-1}$.
The grating, which has 514 lines\,mm$^{-1}$, covers the wavelength range between
\hbox{4800 and 5380\,\AA}. These characteristics imply a sampling of
1.1\,\AA~pixel$^{-1}$ and a spectral resolution of 3.74\,\AA\ (FWHM), which
corresponds to $\sigma_{\rm inst}\sim\,94$\,km\,s$^{-1}$. The detector is an
EEV12 CCD composed of $2148\times4200$ pixels of $13.5\times13.5$ $\mu$m$^2$
each.

\begin{table}
\begin{center}
\caption{Summary of the observations.}
\label{tab:observations}
\begin{tabular}{cccccc}
\hline
Name       & Integration time & Seeing     & Central $S/N$ \\  
(1)        & (2)              & (3)        & (4)           \\
\hline
NGC\,2859  & 4 hours          & 0.9 arcsec & 353           \\
NGC\,3941  & 3.5 hours        & 0.6 arcsec & 389           \\
NGC\,4725  & 3 hours          & 1 arcsec   & 252           \\
NGC\,5850  & 4 hours          & 1.3 arcsec & 161           \\
\hline
\end{tabular}
\end{center}
\begin{minipage}{8.4cm}
(1) Galaxy name; (2) total integration time for each galaxy; (3) mean
integration time-weighted seeing value during the observation 
of the galaxy; (4) maximum $S/N$ value reached for each galaxy. This $S/N$ corresponds
to a spectrum in the central region of the galaxy, where the surface brightness
profile presents also a maximum.
\end{minipage}
\end{table}

The observations were taken with the galaxies centred on the FoV of the lenslet
array. The instrument was orientated so that the major axis of the FoV is
parallel to the major axis of each galaxy disc.
The true orientation of our maps on the sky is indicated in Appendix
\ref{sec:app}.The total integration time for
the galaxies ranges between three and four hours, in exposures of 30 minutes
each. The dithering of 1\,arcsec applied between the different exposures allows
a correction of possible bad pixels of the CCD. Apart from the object lenslets,
\sauron\ has an additional array of 146 lenses for simultaneous sky observations.
These spectra are taken at a distance of 1.9 arcmin from the main FoV. The
instrumental scheme and setup possibilities of \sauron\ are described in
\citet{Baconetal2001}.

At least one Neon lamp exposure was obtained before and after each galaxy
exposure, in order to check the focus of the instrument and to assure an
accurate wavelength calibration. Sky flat fields were taken during the
twilights, whereas Tungsten lamp flat fields and several Lick standard stars of
different spectral types were observed throughout each night. 
The seeing value ranged from 0.5 to 1.6\,arcsec.

\section{Data reduction}
\label{sec:reduction}

The data reduction was performed using the specifically designed {\tt XSAURON}
package, following the prescriptions given in \citet{Baconetal2001} and in
\citet{Emsellemetal2004}. The first steps are the bias subtraction and the
extraction of the spectra. For the latter, it is necessary to create a mask which
indicates which pixels of the CCD correspond to the same spectrum. Such a mask
is characteristic of each observation run. Once the individual spectra are
extracted, they are wavelength-calibrated, flat-field corrected, and cleaned
from cosmic rays. Data for each galaxy are then homogenised by degrading all
the spectra to the same resolution. Then, the spectra are flux-calibrated using
the observed spectrophotometric standard stars. Finally, we align and sum up the
spectra from different exposures, and we create the cubes in the Euro3D format 
\citep{KisslerPatigetal2004}.

\begin{figure*}
\begin{center}
\includegraphics[angle=0,width=.8\linewidth]{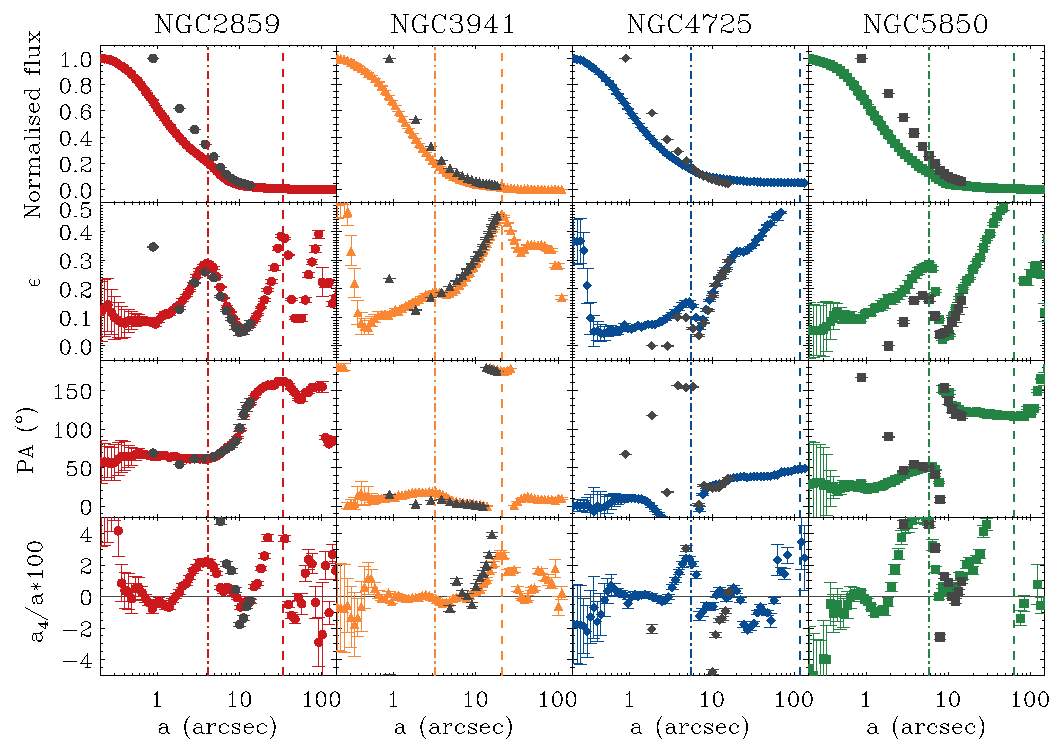}
\caption{From top to bottom: normalised flux, ellipticity, PA, and the fourth
cosine Fourier coefficient ($a_4$) profiles resulting from fitting ellipses to
the isophotes of \emph{i}-band images from SDSS (coloured symbols) and to the
reconstructed \texttt{SAURON} flux maps (overplotted grey symbols) for the four
double-barred galaxies: NGC\,2859, NGC\,3941, NGC\,4725, and NGC\,5850. The
x-axes represent the semi-major lengths of the fitted ellipses. The dot-dashed
and dashed vertical lines indicate the inner and outer bar lengths for each
galaxy, as given by \citet{Erwin2004}.}
\label{fig:morpho}
\end{center}
\end{figure*}

In order to assure the minimum signal-to-noise ratio (hereafter $S/N$) required
for our analysis, we spatially bin the final datacubes using the Voronoi 2D
binning algorithm of \citet{CappellariandCopin2003}. Thus, we create compact
bins with a minimum S/N of $\sim60$ per spectral resolution element, which
allows us to measure the higher order moments of the line-of-sight velocity
distribution (hereafter LOSVD), h$_3$ and h$_4$. Note, however, that the spectra
corresponding to the central regions of the galaxies have already $S/N$ values
well above $60$\,\AA$^{-1}$, so they remain unbinned.\looseness-1

\section{Photometric analysis}
\label{sec:phot}

The relevant bar parameters of the sample have been already analysed by several
authors, mainly \citet{Erwin2004} whose results are compiled in
Table~\ref{tab:sample}. 

In order to assess whether we can \textit{photometrically} resolve the different
structures in our data, we have carried out ellipse fitting over the flux maps
recovered after stacking our spectra in the wavelength direction. For this
purpose we have made use of the {\scriptsize KINEMETRY} program
\citep{Krajnovicetal2006}, which calculates the best fitting ellipses in an
analogous way to how it is done in the IRAF task ELLIPSE \citep{Jedrzejewski87}. The
resulting normalised flux, PA, ellipticity, and the fourth cosine Fourier
coefficient ($a_4$) profiles are shown in Figure~\ref{fig:morpho}, which
confirms the inner and outer bar length estimates given by 
\citet[][as indicated by the corresponding maxima in the 
ellipticity profiles]{Erwin2004}. In the
case of NGC\,3941, the presence of the inner bar is difficult to appreciate in
this photometric analysis due to its small size and ellipticity.
Figure~\ref{fig:morpho} also shows that the $a_4$ parameter tends to peak at the
semi-major axis length of the two bars. This trend was already pointed out by
\citet{deLorenzoCaceresetal2012} for the case of the inner bar of the
double-barred galaxy NGC\,357. The four galaxies presented here are mainly
shaped by discy isophotes ($a_4>0$) especially at the inner and outer bar ends,
with the exception of NGC\,3941, for which the analysis in the inner bar region
indicates pure ellipsoid isophotes.

The results obtained with the \sauron\ data are in good agreement with the
profiles we measured using \emph{i}-band images from the Sloan Digital Sky
Survey \citep[SDSS; ][]{Yorketal2000}. This comparison is shown in
Figure~\ref{fig:morpho}. Deviations occur only at the very central regions, due
to PSF effects. This photometric analysis confirms that our data are well 
suited to the study of our sample of four double-barred galaxies.

\section{Stellar and ionised-gas kinematics}
\label{sec:kin}

\subsection{Stellar kinematics}

To extract the stellar kinematics, we performed a full-spectrum fitting over the
stellar spectra as described in \citet{FalconBarrosoetal2006}. The potential
emission lines included in the \sauron\ spectral range (i.e., H$\beta$,
[OIII]5007 and [NI]) were previously masked in order to work with the pure
absorption contribution. We made use of the penalized pixel fitting (hereafter
{\scriptsize pPXF}) routine developed by \citet{CappellariandEmsellem2004}, which fits
each stellar spectrum with a linear combination of a well-selected set of
templates. In this case, we selected a subsample of the updated single stellar
population models (SSPs) of \citet[][see also
\citealt{FalconBarrosoetal2011}]{Vazdekisetal2010}, with evenly sampled ages and
metallicities. A linear combination of those templates, adjusted to match the
spectral resolution of our data, was convolved with a Gauss-Hermite function
\citep{Gerhard93,vanderMarelandFranx93} to recover the best fitting four
lowest-order moments of the LOSVD (velocity, velocity dispersion, h$_3$, and
h$_4$). The resulting maps are presented in Appendix \ref{sec:app}.
Although the galaxies show no dust content (see Section \ref{sec:sample}), 
the narrow wavelength coverage of \sauron\ additionally
assures that our analysis is not affected by possible differential extinction.

\begin{figure}
\begin{center}
\includegraphics[angle=0,width=\linewidth]{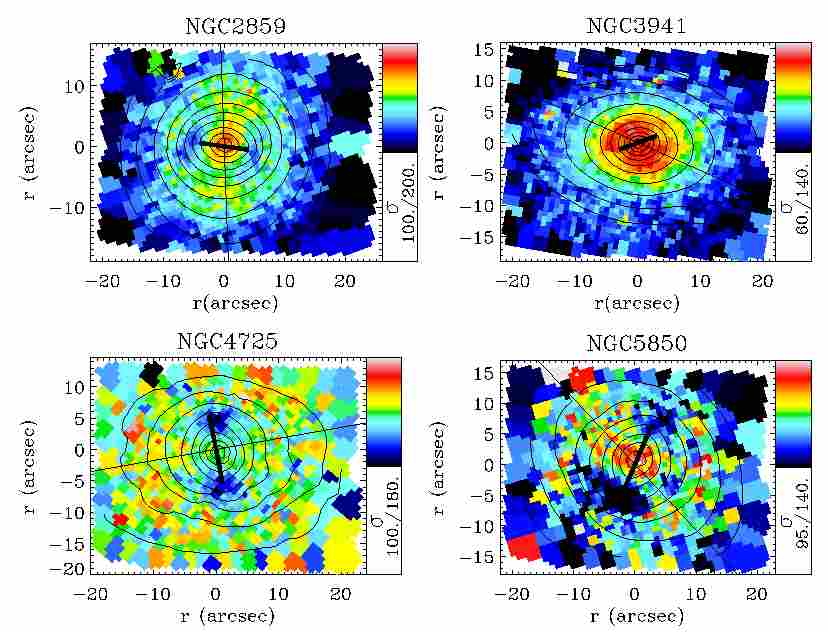}
\caption{Stellar velocity dispersion maps, in km\,s$^{-1}$, for the four
double-barred galaxies: NGC\,2859, NGC\,3941, NGC\,4725, and NGC\,5850. The
length and position angle of the inner and outer bars are indicated by thick
and thin lines, respectively. We have also overplotted the contours of the
reconstructed total intensity maps. The $\sigma$-hollows are clearly seen at the
edges of the four inner bars, as two spots with bluer colours (i.e., lower
values) than the surroundings.}
\label{fig:hollows}
\end{center}
\end{figure}

\begin{figure*}
\begin{center}
\includegraphics[angle=0,width=\linewidth]{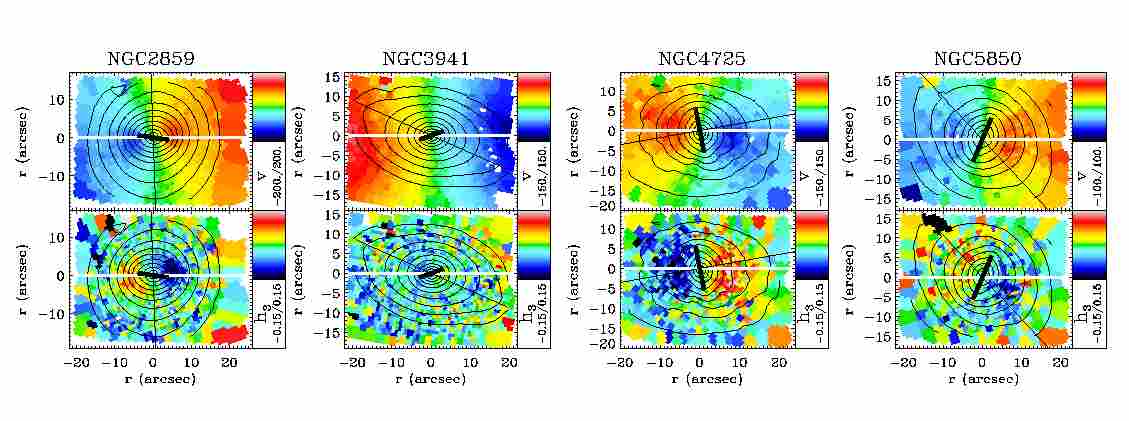}
\caption{Stellar line-of-sight velocity (km\,s$^{-1}$; top panels) and h$_3$
(bottom panels) maps for the double-barred sample. NGC\,2859, NGC\,4725, and
NGC\,5850 show evidence of a kinematically decoupled, corotating inner disc,
seen as two local velocity maxima and minima in the velocity fields at a few
arcseconds from the centres, with an anticorrelated counterpart at the same
locations in the h$_3$ maps. We have overplotted the position angle and length
of the inner (thick line) and outer (thin line) bars, and the contours of the
reconstructed total intensity map, in order to highlight the fact that these
decoupled inner discs are aligned with the semi-major axes of the galaxy discs
(also indicated by a white line)
and thus do not seem to be related to the inner bars.}
\label{fig:discs}
\end{center}
\end{figure*}

\subsubsection{The $\sigma$-hollows}
\label{sec:hollows}

The velocity and velocity dispersion maps for this sample of double-barred
galaxies were already shown in \citet{deLorenzoCaceresetal2008}, where we
presented the discovery of the \emph{$\sigma$-hollows}. The
$\sigma$-hollows are two symmetric, local decreases of the velocity dispersion
values that appear exactly at the edges of the inner bars, with amplitudes with
respect to their immediate surroundings ranging between 10 and 40\,km\,s$^{-1}$,
as seen in Figure \ref{fig:hollows}. In \citet{deLorenzoCaceresetal2008} we
discussed different scenarios that might explain the presence of these hollows,
concluding that they are just a matter of contrast between the $\sigma$ values
of the inner bar and the surrounding bulge. 

The importance of the $\sigma$-hollows lies in that they are a signature of the
inner bars. They are also clearly visible in the velocity dispersion profile of
the double-barred galaxy NGC\,357 \citep{deLorenzoCaceresetal2012}, studied
through long-slit spectroscopy with slits oriented along both the inner and
outer bars. The amplitude of the hollows in NGC\,357 is $\sim$20\,km\,s$^{-1}$
and thus consistent with the values found in the present sample.

The $\sigma$-hollows have not been previously found in other works devoted to
the study of the stellar kinematics of double-barred galaxies, probably due to
the precise requirements needed for their detection. In fact, most existing
observations of double-barred galaxies relied on long-slit spectroscopy along
one direction, typically the major axis of the disc
\citep[e.g.,][]{Corsinietal2003}. Only \citet{Emsellemetal2001}
positioned the slit along the inner bar, but the S/N in their case was not high
enough to clearly distinguish whether the hollows are present in their 
velocity dispersion profiles. Better
suited integral-field spectroscopy was used by \citet{Moiseevetal2004} with the
MPFS instrument. However, their FoV (16$\times$15\,arcsec$^2$) prevented the
detection of the $\sigma$-hollows signature in their sample.

It is interesting to note that none of the galaxies presented in this work shows
a central $\sigma$-drop. The $\sigma$-drop is a local velocity dispersion
minimum that appears at the centre of many spiral galaxies, usually related to a
cold, newly-formed stellar component, such as an inner disc.
\citet{FalconBarrosoetal2006} find this signature in at least 46\% of a sample
of 24 early-type spirals, and other cases have been noticed in different samples
\citep{Marquezetal2003,Gandaetal2006,Comeronetal2008}, especially 
in single-barred galaxies but
also in double-barred galaxies \citep{Emsellemetal2001}. Given the high
frequency of $\sigma$-drops in these kinds of objects, the fact that none of the
four double-barred galaxies shown here presents this signature seems peculiar.

\subsubsection{Kinematically decoupled inner discs}

The stellar velocity maps of the four double-barred galaxies resemble those
expected for unbarred galaxies: they do not seem to be affected by the presence
of the inner nor the outer bars, and just show the rotation along the major axis
of the main galaxy disc. However, for the three cases highlighted in Figure
\ref{fig:discs} (NGC\,2859, NGC\,4725, and NGC\,5850), local velocity
maxima and minima appear between 5 and 10 arcseconds of the galaxy centre, aligned
also with the kinematical major axis. These decoupled structures are rotating
faster than their surroundings and are probably indicating the presence of
stellar inner discs. This hypothesis is supported by the h$_3$ maps also
included in Figure \ref{fig:discs}, which show a clear anticorrelation with
respect to the stellar velocities exactly at the locations of those features, as
expected in the case of discs \citep{BureauandAthanassoula2005}. These features
found in the velocity and h$_3$ maps are unlikely to be related to the inner
bars given their completely different orientations.

\subsection{Gas kinematics}

Besides the stars, our galaxies show emission-line ionised gas, some of which 
fill in the absorption features in the observed spectra. A careful separation of 
this gas from the stars is crucial for a proper analysis of each tracer. Our four 
double-barred galaxies show significant amounts of gas, which we 
recovered by using the {\scriptsize GANDALF} code \citep{Sarzietal2006}. This 
procedure allows us to recover the amplitudes and kinematics of the emission 
lines present in our wavelength range (e.g., H$\beta$, and the [OIII]4959,5007 
and [NI]5198,5200 doublets). Our analysis shows that H$\beta$ and [OIII]5007 are 
the two most prominent lines. While we fitted both lines independently, i.e., 
they were not forced to have the same kinematics, the gas distribution and 
kinematics obtained for each line are very consistent for all galaxies except 
for NGC\,5850, which is discussed in Section \ref{sec:gas5850}.

\begin{figure}
\begin{center}
\includegraphics[angle=0,width=0.48\textwidth]{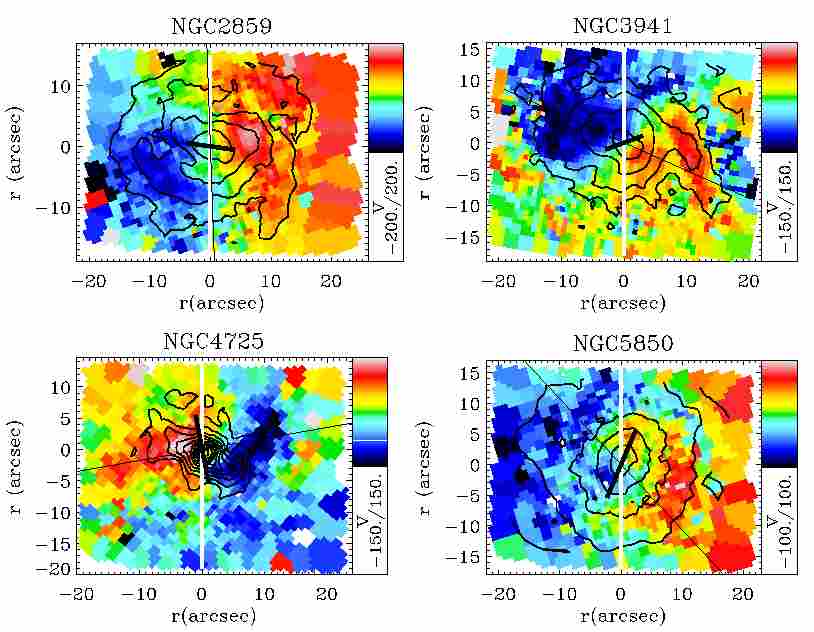}
\caption{Gas velocity maps (km\,s$^{-1}$) corresponding to the [OIII]5007
emission line for the four double-barred galaxies. The zero isovelocity curves
are clearly twisted with respect to the kinematical minor axes. The isodensity
contours of the [OIII]5007 intensities are overplotted in order to emphasize the
distortion of the velocity fields due to their spiral-like gas distributions. The black
lines indicate the position and size of the inner (thick line) and outer (thin
line) bars, whereas the white line shows the direction of the kinematical minor axes.}
  \label{fig:gas}
\end{center}
\end{figure}

\subsubsection{Possible evidence of gas inflow to the central region}

The gas intensity maps for the four double-barred galaxies reveal interesting
spiral structures, especially evident 
in the cases of NGC\,2859 and NGC\,3941.
Moreover, the velocity maps also seem to be disturbed by these spirals, as their
zero isovelocity curves appear twisted with respect to the kinematical minor
axes. Figure~\ref{fig:gas} highlights the distortion of the velocity fields in
comparison with the gas intensities. The spiral structures and the twisting of
the velocity fields suggest there are gas streaming motions towards the central
region of these double-barred galaxies \citep{Fathietal2006}. 
This is most notable in the case of
NGC\,5850.

In general, the gas distribution of our galaxies does not seem to be related to 
the main bars. This behaviour is unexpected, as ionised gas is commonly observed 
tracing the often present dust lanes seen along the main bar major axis 
\citep[e.g., ][]{Athanassoula92,PatsisandAthanassoula2000}. Our result may 
be connected to the early-type nature of our galaxies (except for NGC\,5850), 
which means that they do not have an appreciable dust content,
as well as to the limited FoV sampled by our data. Unlike the case of the stellar 
kinematics, the gas velocity dispersion maps are quite flat and show no 
signatures of the presence of the inner or the outer bars 
(see Appendix \ref{sec:app}).

\subsubsection{H$\beta$ vs. [OIII]5007 gas distributions in NGC\,5850}
\label{sec:gas5850}

\begin{figure}
\begin{center}
\includegraphics[width=0.65\linewidth]{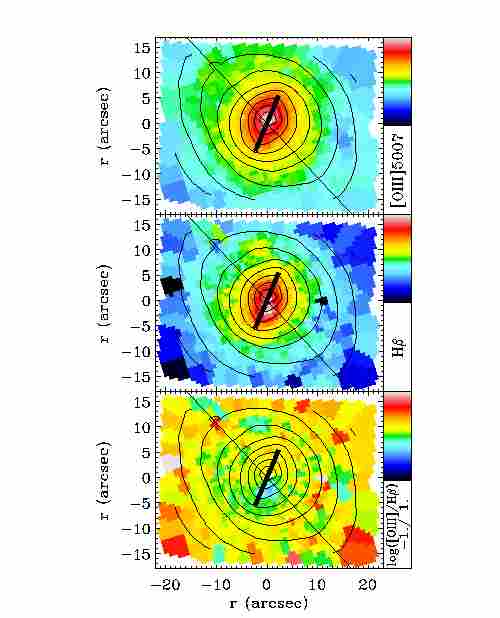}
\caption{Gas intensity maps for NGC\,5850. The top panel shows the intensity
distribution of the [OIII]5007 emission line, whereas the middle panel corresponds to 
the H$\beta$ intensity. The distributions
are different, with the H$\beta$ peak off-centred with
respect to the stellar (shown with contours) and [OIII] intensities. The
corresponding kinematic maps (not shown here) are however equivalent, within the
resolution of our data. The bottom panel shows the [OIII]/H$\beta$ ratio,
which seems to indicate the presence of a small star-forming region at the
position of the maximum H$\beta$ emission. For reference, the straight lines
indicate the position and size of the inner (thick line) and outer (thin line)
bars.}
\label{fig:gas5850}
\end{center}
\end{figure}

NGC\,5850 is the only galaxy for which the gas distributions derived from the
[OIII]5007 and H$\beta$ emission lines are different. The corresponding maps are
shown in Figure \ref{fig:gas5850}. Whereas the [OIII]5007 distribution matches
well with the stellar intensity distribution, the H$\beta$ map shows the maximum
off-centred with respect to the stellar isophotes. However, this remarkable
difference is not translated into kinematical differences, at least within the
resolution of our data, since the gas velocity and velocity dispersion maps
obtained from the two emission lines seem equivalent.

The position of the H$\beta$ emission peak in NGC\,5850 corresponds with a small
potential star-forming region, as indicated by the [OIII]/H$\beta$ ratio map
also shown in Figure \ref{fig:gas5850}. This result is confirmed by
\citet{Bremeretal2012} in a study of optical diagnostic diagrams using \texttt{VIMOS}
integral-field spectroscopy. In addition they found a faint, second star-forming
region 1.6\,kpc (i.e., $\sim$11.4 arcsec) East of the centre, which is
not observed in our data. Their study favours the LINER nature of the nucleus. 
Note also that none of these potential
star-forming regions is spatially coincident with the CO peak
mentioned in Section \ref{sec:sample} \citep{Leonetal2000}, which
is located to the North of the centre.
Unfortunately our data do not allow us to
determine the extent to which the star-forming region(s) are due to the inner
bar.

\begin{figure}
\begin{center}
\includegraphics[angle=0,width=0.6\linewidth]{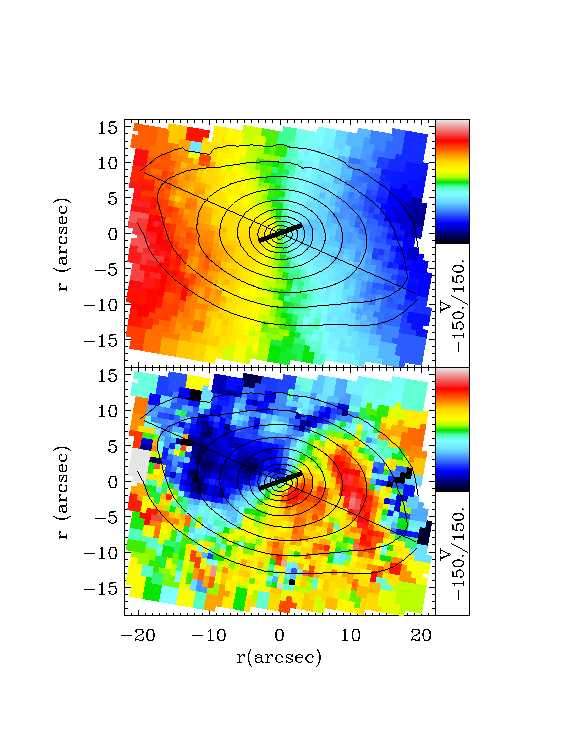}
\caption{Stellar (top panel) and gas (bottom panel) velocity maps for NGC\,3941,
in km\,s$^{-1}$. The gas in NGC\,3941 is clearly counter-rotating with respect
to the stars. For reference, we have overplotted the position angle of the inner
(thick line) and outer bar (thin line), and the contours of the
reconstructed total intensity map.}
\label{fig:kin3941}
\end{center}
\end{figure}
\begin{figure*}
\begin{center}
\includegraphics[width=\linewidth]{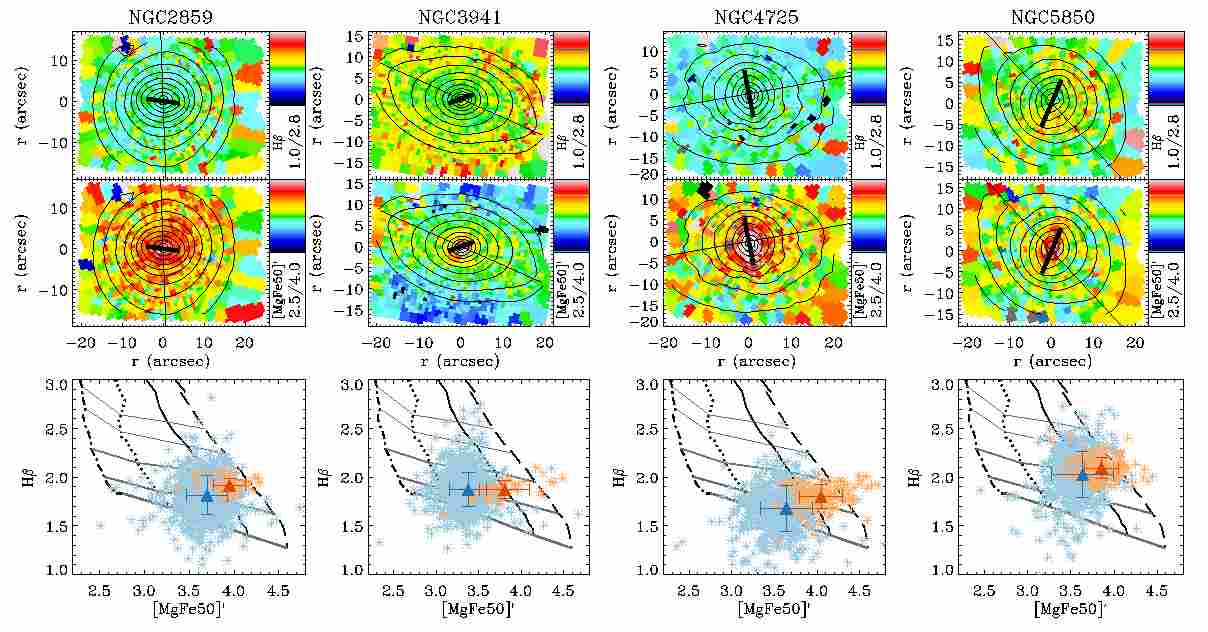}
\caption{H$\beta$ (top panels) and [MgFe50]$^\prime$ (middle panels) index
distributions for NGC\,2859, NGC\,3941, NGC\,4725, and NGC\,5850 (from left to
right). The slightly higher values of [MgFe50]$^\prime$ present in the inner bar
regions of the four double-barred galaxies suggest they are more metal-rich than
their surroundings. The length and position angle of the inner and outer bars
are indicated by thick and thin lines, respectively, and the contours of the
reconstructed total intensity maps are also overplotted. The bottom row shows the
H$\beta$ versus [MgFe50]$^\prime$ measurements corresponding to the SSP models by
\citet{Vazdekisetal2010}. The solid lines represent different ages increasing
from top to bottom (2.5, 3, 5.6, 10, and 18\,Gyr, respectively), whereas the
almost vertical lines indicate different metallicities increasing from left to
right ([Z/H]=\,$-0.7$, $-0.4$, 0.0, and 0.2, respectively). The asterisks are
the measurements for all the Voronoi bins of each galaxy, distinguishing
the bins corresponding to the inner bar (orange asterisks) from those
corresponding to the outer bar (blue asterisks). The triangles represent the
average measurements for the inner bar (red triangles), and the rest of the
galaxy (blue triangles). The average values indicate that inner bars are coeval
or slightly younger than their surroundings. Both models and galaxy measurements 
are carried out in the LIS-8.4\,\AA\ system.}
\label{fig:indices}
\end{center}
\end{figure*}

\subsubsection{Counter-rotating gas in NGC\,3941}

Figure \ref{fig:kin3941} compares the velocity fields corresponding to the stars
and gas in NGC\,3941, showing that the components are clearly decoupled and
almost counter-rotating. Counter-rotation is not a peculiar feature, as
\citet{Bertolaetal92} estimated that 10\% to 20\% of S0 galaxies have such a
gas component. Subsequent studies have found even higher values ($\gtrsim$\,20\%)
for this fraction \citep[e.g.,][among
others]{Pizzellaetal2004,BureauandChung2006}. These results contrast with the
markedly different incidence of counter-rotation in stellar components
($\sim$10\% in S0s, see \citealt{Kuijkenetal96,Emsellemetal2011}). It is
therefore worth noting cases where gaseous and stellar counter-rotating discs
are present (see the case of NGC\,5719 in \citealt{Coccatoetal2011}).
 
The counter-rotation in NGC\,3941 was already noticed by \citet{Fisher97}, who
suggests that it is probably the result of a merger event or accretion of gas
with opposite angular momentum between them. In fact, merging is 
one of the most popular
explanations for the presence of these kinds of structures
\citep{Cirietal95,Barnes2002}, even in the case of stellar discs
\citep{ElicheMoraletal2011}. There are however other possibilities, such as the
appearance of \emph{anomalous orbits} within a triaxial and tumbling potential,
such as that of a bar \citep{EmsellemandArsenault97,FalconBarrosoetal2004}. These
anomalous orbits are a family of close, stable, retrograde orbits, which are
tilted with respect to the equatorial plane. Therefore, they can capture the
ionised gas, thus forming the counter-rotating disc. Given that the kinematical
major axis of the gas in NGC\,3941 seems to be aligned with the main bar and that this
galaxy is isolated, anomalous orbits seem to be the most likely explanation for the
counter-rotating gas disc in NGC\,3941.

\section{Stellar populations}
\label{sec:sp}

\subsection{Line-strength maps}
\label{sec:linemaps}

In order to investigate which stellar populations are shaping the double-barred
galaxies, we measured the most relevant absorption line-strength indices over the
emission-cleaned spectra covering the full \sauron\ FoV. The first necessary
step, for the proper comparison of all Voronoi bins and galaxies, was the
homogenisation (i.e., degradation) of all spectra to the same total broadening.
We chose to degrade our data to a total full-width half maximum of 8.4\,\AA\ to
match the LIS-8.4\,\AA\ system proposed by \citet{Vazdekisetal2010}. This new
LIS system introduces several advantages with respect to the traditional
Lick/IDS spectral system \citep{Wortheyetal94}: it avoids the lack of
flux-calibration and the well-known wavelength dependence of the resolution
\citep{WortheyandOttaviani97}.

We calculated the age-sensitive H$\beta$ and the metallicity-sensitive Mg\emph{b}
and Fe5015 indices, following the Lick definitions given by \citet{Trageretal98}. 
Total metallicity is estimated from the [MgFe50]$'$\footnote{$\rm [{\rm
MgFe50}]^\prime = \frac{0.69 \times {\rm Mg}\emph{b} + Fe5015}{2}$} index, defined by
\citet{Kuntschneretal2010}. [MgFe50]$^\prime$ has been proved to be a good metallicity
indicator, almost insensitive to the [Mg/Fe] overabundance. This index is mostly
used by users of the \sauron\ spectrograph \citep[see for
example][]{Gandaetal2007}, since it does not require the measurements of the
Fe5270 and Fe5335 indices, necessary to compute the more common [MgFe]$^\prime$
index \citep{Thomasetal2003}, which lie outside of the \sauron\ spectral range.

Figure~\ref{fig:indices} shows the H$\beta$ and [MgFe50]$^\prime$ line-strength
index distributions for the double-barred sample. These maps, together with
those corresponding to the Mg\emph{b} and Fe5015 indices, are
included in Appendix~\ref{sec:app} for each galaxy. It is remarkable that the
total metallicity indicator [MgFe50]$^\prime$ presents slightly higher values at
the regions delimited by the inner bars. This fact suggests that inner bars are
more metal-rich than their surroundings. 

\subsection{Quantifying the stellar population parameters}
\label{sec:ageandmet}

Double-barred galaxies are made of several components and, while 
a scenario in which the different structures are shaped by the
redistribution of the existing stars formed in a single burst is possible,
each region most likely results from the superposition 
of several stellar populations.
Assuming that our galaxies are well described by SSPs, 
it is possible to derive the luminosity-weighted age and metallicity of
all bins by comparing the line-strength measurements with the corresponding
predictions from the SSP models. As for the kinematics, in the stellar
population analysis we use the stellar population models of
\citet{Vazdekisetal2010}. The spectral resolution of these models is
2.51\,\AA\,(FWHM), constant over the whole spectral range
\citep{FalconBarrosoetal2011}. In order to compare the model predictions to the
data, the model spectra have also been smoothed to the LIS-8.4\,\AA\ system.
Figure~\ref{fig:indices} (bottom row) shows the results for the H$\beta$ versus
[MgFe50]$^\prime$ indices. The grids are not perfectly orthogonal due to the
age-metallicity degeneracy. For simplicity, we distinguish between the
regions dominated by the inner bar and its surroundings, i.e., mostly the outer
bar. Similar plots for the rest of the metallicity-sensitive indices can be seen
in Appendix~\ref{sec:appgrids}. 

\begin{figure*}
\begin{center}
\includegraphics[width=\textwidth]{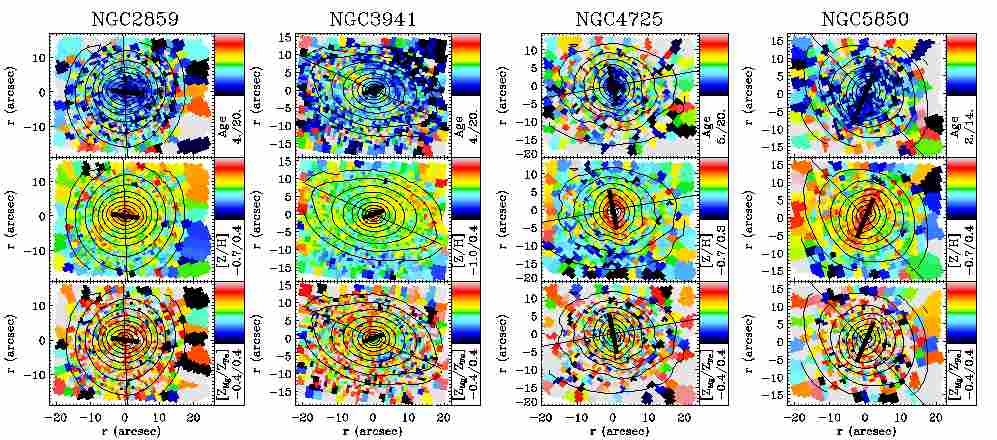}
\caption{Luminosity-weighted age (in Gyr; top panels), metallicity ([Z/H];
middle panels), and [Z$_{\rm Mg}$/Z$_{\rm Fe}$] (bottom panels) maps for
NGC\,2859, NGC\,3941, NGC\,4725, and NGC\,5850 (from left to right,
respectively). The maps are the result of interpolating the H$\beta$ versus
[MgFe50]$^\prime$ grids using the {\scriptsize rmodel} package. For the sake of
clarity, we have overplotted the position angle and length of the inner (thick
line) and outer (thin line) bars, and the contours of the reconstructed total
intensity map. The inner bars appear clearly younger and more metal-rich than
their surroundings, whereas the [Z$_{\rm Mg}$/Z$_{\rm Fe}$] maps show rather flat
distributions.}
\label{fig:ageandmet}
\end{center}
\end{figure*}

Since double-barred galaxies are structurally complex and the two-dimensional
information is crucial, we created age and metallicity maps to investigate the
distribution of the stellar populations. For this purpose, we adapted the {\scriptsize
rmodel}\footnote{www.ucm.es/info/Astrof/software/rmodel/rmodel.html} code
\citep{Cardieletal2003}, specifically developed for this kind of stellar
population analysis, to perform a bivariate interpolation of the H$\beta$ versus
[MgFe50]$^\prime$ grids. The resulting luminosity-weighted age and metallicity
maps are shown in Figure~\ref{fig:ageandmet}. Interestingly, as seen in the
bottom panels of Figure~\ref{fig:indices}, the four inner bars appear not only
more metal-rich, but also slightly younger, than their surroundings. 

[Mg/Fe], [CN/Fe], and other similar abundance ratios are fundamental parameters
in a stellar population analysis, as they act as chemical clocks and allow us to
infer the formation timescales of the bulk of the stellar population. In an
analogous way to the [MgFe50]$^\prime$ index, it is possible to derive the
Mg\emph{b} and Fe5015 metallicity distributions from the interpolation of the
corresponding H$\beta$ versus Mg\emph{b} and H$\beta$ versus Fe5015 model grids.
The [Z$_{{\rm Mg}b}$/Z$_{\rm Fe5015}$] ratio distribution can be then obtained
in a heuristic way by subtracting the two metallicity maps. [Z$_{{\rm
Mg}b}$/Z$_{\rm Fe5015}$] can be considered a proxy of the [Mg/Fe] overabundance,
as a linear relation between the two ratios is expected
\citep{Peletieretal2007,Vazdekisetal2010}; note however that the absolute values
of the [Z$_{\rm Mg}$/Z$_{\rm Fe}$] measurements are not the same as those used
by other authors, who usually take combined Fe-indices, such as
Fe3\footnote{${\rm Fe3=\frac{Fe4383+Fe5270+Fe5335}{3.}}$}
\citep[e.g.,][]{Thomasetal2011,delaRosaetal2011}. It is for that reason, and in
order to avoid confusion with values in the literature, that we refer to our [Mg/Fe]
estimate as [Z$_{\rm Mg}$/Z$_{\rm Fe}$].

Given the small dynamical range of the [Mg/Fe] parameter, which typically
acquires values between 0.0 (i.e., solar abundance) and 0.4, these kinds of maps
are naturally noisier than the age or metallicity maps. The final [Z$_{\rm
Mg}$/Z$_{\rm Fe}$] overabundance distributions for the four double-barred
galaxies are shown in Figure \ref{fig:ageandmet}. They present flat
distributions with typical values above 0. We therefore do not find significant
differences between the inner and outer bar regions. 

\subsection{Radial profiles of the stellar population properties}
\label{sec:radial}

\begin{figure}
\begin{center}
\includegraphics[angle=0,width=0.5\textwidth]{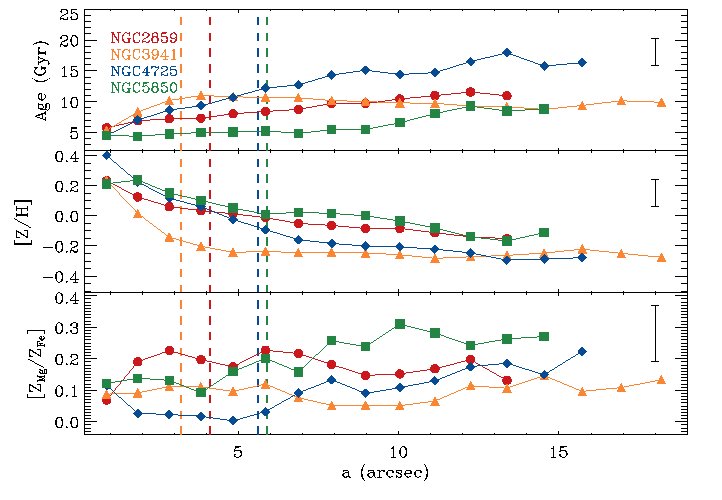}
\caption{Mean luminosity-weighted age (top panel), metallicity (middle panel),
and [Z$_{\rm Mg}$/Z$_{\rm Fe}$] (bottom panel) profiles measured over the best
fitting ellipses of the corresponding intensity map (see text for details). The
x-axis represents the semi-major axes of the ellipses. Different colours and
symbols correspond to different galaxies: NGC\,2859 (red circles); NGC\,3941
(yellow triangles), NGC\,4725 (blue diamonds), and NGC\,5850 (green squares).
Vertical dashed lines indicate the length of the inner bars, as given in
Table~\ref{tab:sample}, whereas the mean error is indicated at the top right of
each panel. The positive and negative gradients outwards in age and metallicity,
respectively, are evident, whereas the [Z$_{\rm Mg}$/Z$_{\rm Fe}$] profile shows
a slight trend towards lower values at the central regions.}
\label{fig:radial}
\end{center}
\end{figure}

In order to study in more detail the behaviour of the stellar population
parameters from the inner to the outer regions in our data, we have also
extracted radial profiles of the luminosity-weighted age, metallicity, and
[Z$_{\rm Mg}$/Z$_{\rm Fe}$]. For this purpose we recover the PA, ellipticity,
and semi-major axes of the best fitting ellipses to the flux maps calculated in
Section~\ref{sec:phot}, and run {\scriptsize KINEMETRY} again over the age,
metallicity, and [Z$_{\rm Mg}$/Z$_{\rm Fe}$] maps of Figure~\ref{fig:ageandmet}
to get the mean values of those parameters over the isophotes.
Figure~\ref{fig:radial} shows that three out of the four double-barred galaxies
present positive age gradients and negative metallicity gradients going
outwards. The exception to this behaviour is illustrated by NGC\,3941, which
shows flat gradients in age, metallicity and abundance ratio in the regions 
dominated by the main bar.

\citet{Perezetal2009} obtained the mean age and metallicity gradients along the
main bars for a sample of 20 barred galaxies, eight of which are double-barred
systems with one (NGC\,2859) in common with our present sample. They found
that main bars show both positive and negative age and metallicity gradients
with no correlation between the two parameters, although there is a trend so
that bars with negative metallicity gradients usually show positive age
gradients, as in the general case of the galaxies presented here. Even among the
double-barred galaxies of the \citet{Perezetal2009} sample there are no clear
correlations, with 50\% of their double-bars showing positive age gradients, and
also 50\% positive metallicity gradients. 

The variety of behaviours obtained by \citet{Perezetal2009} is in contrast with 
the homogeneous trend found for our sample. However, \citet{Perezetal2009}
find that negative metallicity gradients tend to appear in galaxies with the lowest central stellar 
velocity dispersions, with $\sigma$\,$<$\,170\,km\,s$^{-1}$. Given that 
our four double-barred galaxies have maximum
central velocity dispersion values ranging between 140 and 170 km\,s$^{-1}$, our results
are actually in agreement with those of \citet{Perezetal2009}. Moreover,
the radial profiles shown by \citet{Perezetal2009} for NGC\,2859 are
fully consistent with those of Figure~\ref{fig:radial}.

The [Z$_{\rm Mg}$/Z$_{\rm Fe}$] radial profiles are rather flat with
values ranging between 0.0 (i.e., solar abundance) and 0.3 dex, also in
agreement with the results of \citet{Perezetal2009}. All the galaxies show a
slight trend towards a positive gradient, so the central regions present lower
values of the [Z$_{\rm Mg}$/Z$_{\rm Fe}$] overabundance. This feature is almost
unnoticed in the [Z$_{\rm Mg}$/Z$_{\rm Fe}$] maps, which stresses the importance of
analysing the data in different ways. The implications of this finding will be
discussed in Section~\ref{sec:discussion}.

\subsection{The characteristic stellar population properties of each structural component}
\label{sec:integrated}

Our last approach to analyse the stellar population content of the double-barred
galaxies is to obtain the mean luminosity-weighted age, metallicity, and
[Z$_{\rm Mg}$/Z$_{\rm Fe}$] values in three regions of interest: a central
region defined as a circle of radius 1\,arcsec; the inner bar, defined as an
ellipse with the corresponding PA, ellipticity, and semi-major axis
characteristic of each inner bar (as given in Table \ref{tab:sample}),
after subtracting the central 1\,arcsec-circle; and the
outer bar region, which is defined as an ellipse with the PA and ellipticity of
the corresponding outer bar as given in Table \ref{tab:sample} and a semi-major
axis that is 5\,arcsec longer than the semi-major axis length of the inner bar.
With this criterion we assure that the considered outer bar region is
inside the \texttt{SAURON} FoV and we avoid the very outer bins that present
lower $S/N$. The central bins corresponding to the inner bar region are also removed
from the outer bar definition in order to avoid the mixing of components. 
We deliberately ignore the bulge as it is not straightforward 
to delimit its extent from our data. 

\begin{figure}
\begin{center}
\includegraphics[angle=0,width=0.5\textwidth]{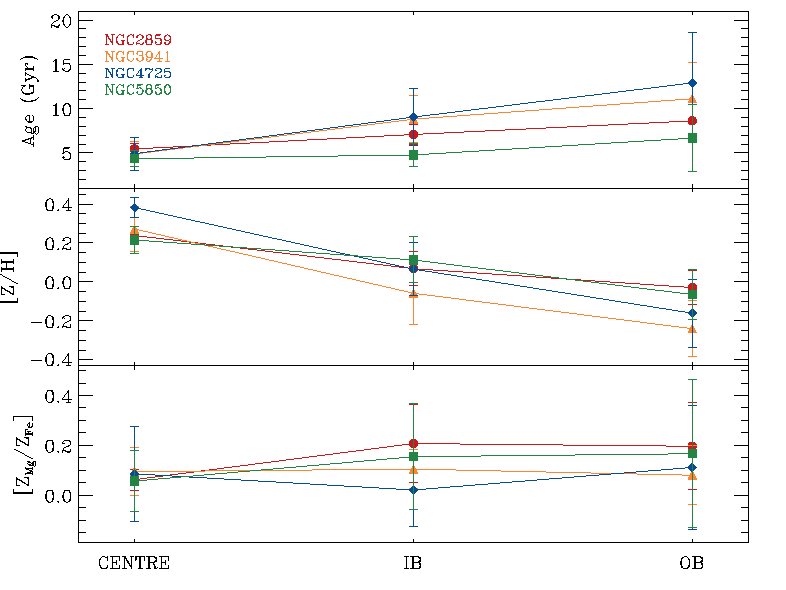}
\caption{Mean luminosity-weighted age (top panel), metallicity (middle panel),
and [Z$_{\rm Mg}$/Z$_{\rm Fe}$] (bottom panel) values for the regions
corresponding to a central circle with a 1\,arcsec radius (CENTRE), inner
bar (IB), and outer bar (OB; see text for details on the OB definition). 
Different colours and symbols
correspond to different galaxies: NGC\,2859 (red circles), NGC\,3941 (yellow
triangles), NGC\,4725 (blue diamonds), and NGC\,5850 (green squares). The error
bars indicate the dispersion of the values inside each region. The positive and
negative gradients outwards in age and metallicity, respectively, are evident.
The [Z$_{\rm Mg}$/Z$_{\rm Fe}$] distribution shows a slight trend towards lower
values at the centre.}
\label{fig:integrated}
\end{center}
\end{figure}

Figure~\ref{fig:integrated} shows the mean values of the parameters
in the bins inside each region of interest, with the corresponding dispersion
represented by the error bars. Note that, as expected, the general trends shown in 
Section~\ref{sec:radial} appear also among structural components, namely the inner
structures are always younger and more metal-rich than the outer ones, as seen in
the age and metallicity maps of Figure~\ref{fig:ageandmet}. 

\citet{deLorenzoCaceresetal2012} studied the stellar populations of the bulge,
inner bar, and outer bar of the double-barred early-type galaxy NGC\,357. 
They found that inner structures, i.e., bulge and inner bar, showed
approximately the same mean luminosity-weighted age ($\sim$8 Gyr) and solar
metallicity, whereas the outer bar had a similar age and clearly lower
metallicity. Taking into account that \citet{deLorenzoCaceresetal2012}
integrated the whole bulge region and not only the very central part, the results
obtained for NGC\,357 are compatible with the general conclusions derived for
the present sample. In fact, the negative metallicity gradient exists in the
five double-barred galaxies. Moreover, the five inner bars show approximately
solar metallicity values and ages around $\sim$7\,Gyr.

The [Z$_{\rm Mg}$/Z$_{\rm Fe}$] ratios for the central, inner bar, and outer bar
regions also show the previously observed trend towards lower values in the
central structures. Note that the four galaxies have a central abundance ratio
around solar value, with [Z$_{\rm Mg}$/Z$_{\rm Fe}$]$\sim0.05$\,dex. It is
known that ellipticals tend to have enhanced abundance ratios
\citep[e.g.,][]{Wortheyetal92}, and bulges of many early-type spirals usually
mimic the $\alpha$-enhancement of ellipticals with a similar velocity
dispersion
\citep[e.g.,][]{Jorgensen99,Kuntschner2000,Vazdekisetal2004,Carreteroetal2004}.
Following this relation, early-type spirals of central stellar velocity
dispersions of 140--180 km\,s$^{-1}$ should have [Mg/Fe]$\sim0.2$-$0.4$\,dex,
well above the central values found for our double-barred sample and closer to the
mean abundance ratios of the bars. The low values presented here are nonetheless 
not unexpected if some recent star formation has taken place in the inner regions
of the galaxies \citep[e.g., ][]{ThomasDavies06}.

\section{Discussion}
\label{sec:discussion}

\subsection{The formation and stability of double-barred systems}

The formation of double-barred galaxies has been a matter of debate for more
than two decades. Most of the work done to address this issue is based on
numerical simulations that usually require the presence of a dissipative
component, i.e., gas content, which is dynamically disturbed to form
the inner bar. \citet{FriedliandMartinet93} tried to develop double-barred
systems with numerical simulations including only a purely collisionless
component. They concluded that the presence of the gas was needed to
form the two bars. Once gas was included, they got different models in which the
two bars formed simultaneously, or in which the inner bar was formed after (and
thanks to) the main bar. These two bars rapidly decoupled from each other and
rotated with different pattern speeds. This last result has been observationally
confirmed \citep{Corsinietal2003}. Those initial simulations predicted, however, that
the final systems were not long-lived since at least the inner bar would
disappear 1--2\,Gyr after its formation.

Since 1993, many more studies have succeeded at simulating the formation
\citep{Helleretal2001,ShlosmanandHeller2002,Rautiainenetal2002,
EnglmaierandShlosman2004,Helleretal2007} and also the dynamical evolution 
\citep[e.g.,][]{MaciejewskiandSparke2000,MaciejewskiandAthanassoula2007} of 
double-barred systems. These simulations have in common that they also need
a fundamental dissipative component to create inner bars, mostly because gas flows
along the outer bar and is captured by its $x_2$ orbits, subsequently 
decoupling and forming the inner bar.
Within these scenarios, inner bars should be younger and probably more metal-rich 
than their surroundings.

The observational results presented in this work are compatible
with the bulk of simulations that need dissipation to form the inner bars. 
The analysis of the emission lines reveals that the whole sample actually contains ionised gas
and that this is probably flowing towards the central regions. Moreover, our
inner bars have turned out to be slightly younger and more metal-rich than their
surroundings (i.e., the outer bar). The gas-rich scenario is 
further supported by the ubiquitous presence of 
extended HI gas in all the galaxies of our sample.

In contrast to the simulations mentioned above, \citet{DebattistaandShen2007}
generated long-lived inner bars without the presence of any dissipative component,
although they require the presence of a rapidly rotating structure
at the centre of the galaxy that 
is created \emph{ad hoc} in these simulations.
Inner bars then formed through the redistribution of the already existing
stars. Therefore, at the moment of bar formation there are no
expected differences among the stellar populations of the inner bars and the surrounding
components, unless some additional gas were present or accreted into the
central regions. While such a burst might lead to a small age difference, it is
not clear whether it could also reproduce the observed metallicity gradient.

Note that for our analysis we only use single stellar populations
leading to mean luminosity-weighted ages, which are biased towards the most
recent stellar components. We find a significant but not large difference
between the derived ages for the inner and outer bars. 
It is important to be aware of the fact that in order to observationally disentangle 
the formation sequence of inner bars, not only the star formation
history but also the dynamical evolution of the different structures
need to be known. In fact, inner
bars are dynamically distinguished structures and their formation might have a dynamical origin with
no effect over their stellar populations. Recovering both the star formation history
\emph{and} dynamical evolution of a structurally complex system such as a double-barred galaxy
is a very challenging goal.

Regarding the stability of bars, it is worth noting that our inner bars are not particularly young
systems in absolute terms, since they have mean luminosity-weighted ages ranging from 4 to 8 Gyr. 
If the age of the stellar populations traces the age of the stellar inner bar structure,
these intermediate-to-old mean luminosity-weighted values seem to indicate
that inner bars are long-lived systems, in agreement with most of the
simulations and with the high frequency of double-barred galaxies found in the
Universe \citep{ErwinandSparke2002,Laineetal2002}. Disentangling the dynamical evolution of inner bars
is again mandatory to take final conclusions on their stability.

\subsection{The role of bars in galaxy evolution}

Bars are non-axisymmetric components that redistribute the angular momentum of a
galaxy, allowing the flow of material along them. It is therefore natural that
inner bars are thought to be key systems in transporting gas to the very central
regions, where they help to trigger star formation and contribute to the
creation of spheroidal structures. Under this framework, double-barred galaxies would be
outstanding systems from a secular evolution point of view, since they would
promote the formation of bulges 
\citep[e.g.,][]{KormendyandKennicutt2004,Athanassoula2005}. Furthermore, it has been theoretically
demonstrated that inner bars allow the material to get to the innermost regions,
where the gas flowing along a single bar is not able to reach. Within this
scenario, inner bars might also contribute to the fueling of active galactic
nuclei \citep[hereafter AGNs;][]{Shlosmanetal89,Shlosmanetal90}. 

In our analysis, we do not find differences between the galaxies hosting
or not an AGN, so we are not able to shed light on the role of inner bars in fueling the AGNs.
As indicated in Section \ref{sec:sample}, NGC\,3941 and NGC\,4725 are Seyfert 2 galaxies
\citep{Veron-Cettyetal2006}, NGC\,5850 hosts a LINER \citep[][]{Bremeretal2012}, 
and NGC\,2859 shows no signs of
nuclear activity. More double-barred galaxies are needed to 
do a proper statistical analysis and to draw conclusions on this matter.

The presence of ionised gas in the four observed double-barred galaxies and the 
rather young ages
of the inner bars suggest that gas has actually reached the inner structures,
more likely by flowing along outer bars, and triggered star formation there.
These results therefore support the idea that secular evolution is taking place in the
double-barred sample, at least at outer bar scales, and that it has had a major effect 
over the stellar population composition of inner bars.

Signatures of secular evolution also appear at inner bar scales, since we have also found possible 
evidence of gas flow through the inner bars
towards the very central regions. It is therefore probable that some low-level star formation is taking place
in these innermost regions.
This hypothesis is also backed by the fact that the centres of our four double-barred galaxies
present somewhat lower [Mg/Fe] values than the inner bars and outer regions, which
points towards a progressive, time-extended formation.
Notwithstanding, this star formation seems to be not efficient nor strong enough to 
be inducing major morphological changes in the four double-barred galaxies. 
If inner bars were playing a major role in the formation of new
structures in the central regions, these new structures should show up in a stellar population
analysis such as the one performed in this work.

\section{Summary and Conclusions}\label{sec:conclusions}
We have performed a detailed analysis of the kinematics and stellar populations
of a sample of four double-barred early-type galaxies
(NGC\,2859, NGC\,3941, NGC\,4725, and NGC\,5850), paying special attention 
to the spatial distribution of the properties provided by the high-quality
\texttt{SAURON} integral-field spectroscopic data. 

The analysis of the stellar and gas kinematics reveals a wide variety of structures:
three of the four galaxies show signatures of the presence of a kinematically-decoupled
stellar inner disc, which supports
the idea that most barred galaxies contain a disc-like central component
\citep{Perezetal2009}, although it is important to notice that none of our four
double-barred galaxies present a $\sigma$-drop at their centres \citep{Emsellemetal2001},
despite the fact that these drops usually appear in barred galaxies.

The stellar $\sigma$-hollows \citep{deLorenzoCaceresetal2008} appear in the four galaxies exactly 
at the edges of the inner bars. These hollows are kinematical signatures of the presence
of inner bars, produced by the contrast between the high stellar velocity dispersion of the
bulge and the lower velocity dispersion of the bar.

Concerning the gas kinematics, we find possible
evidence of gas inflow towards the central regions
in both the gas intensity and velocity maps. One out of the four galaxies, NGC\,3941, contains
gas which is counter-rotating with respect to the stellar component, whereas NGC\,5850
is the only galaxy which shows differences between the [OIII]5007 and H$\beta$
gas distributions.

The main result of this work is the distinct stellar populations found for the inner
bars of the four galaxies: they are clearly younger and more metal-rich than the outer bars.
Moreover, we find positive age and negative metallicity gradients along the inner and
outer bars for all the galaxies except NGC\,3941, which only shows the gradients inside the inner
bar region, whereas the profiles along the outer bars are flat. It is important to 
notice that NGC\,3941 tends to stand out with respect to the other three galaxies in 
all the studies: its inner bar is the smallest and most difficult to appreciate in the 
photometrical analysis and in the maps, it does not contain an inner disc, its gas is counter-rotating,
and its stellar population radial profiles present slight differences with respect to the others.
Notwithstanding, the main conclusions of this work apply also to this galaxy.

We have also analysed the [Mg/Fe] abundance ratio for the four double-barred galaxies, 
finding rather flat overabundance distributions; 
however, the very central regions of the four galaxies present almost solar
[Z$_{\rm Mg}$/Z$_{\rm Fe}$] ratios, with values slightly below the mean overabundances
of the inner and outer bars. This result indicates a long star formation process 
in the very central region of these galaxies.

Although some of the results obtained in this work seem to indicate that inner bars are contributing
to star formation at the centres of the galaxies, the efficiency of this process 
is, at present, low and thus not important enough to promote major morphological changes. 
This conclusion is somewhat understandable for our double-barred galaxies, since they are
mainly early-type objects, for which the secular processes are expected to play a moderate role
in the morphological evolution.

\section*{Acknowledgments}
The authors are indebted to Isaac Shlosman, Reynier Peletier, John Beckman, Jairo M\'endez-Abreu,
and J. Alfonso L. Aguerri for many fruitful discussions and comments on the manuscript.
The careful reading as well as the interesting comments from the anonymous referee have
substantially improved the article. 
Support from the \sauron\ team has been essential during the observations and
early stages of the data analysis.
AdLC thanks the \emph{barred galaxies-girls} 
(Patricia S\'anchez-Bl\'azquez, Isabel P\'erez, and Inma Mart\'inez-Valpuesta)
for their sincere interest in this work, Agnieszka Rys for her advice about the writing,
and Luis Peralta for his positive attitude. AdLC
also acknowledges partial financial support from ESA/ESTEC and a European Union
EARA Fellowship for visiting the Leiden Observatory.
JFB is supported by a Ram\'on y Cajal fellowship of the Spanish Ministry of Economy and Competitiveness.
This research has been funded by the
Spanish Ministry of Economy and Competitiveness 
under the grants AYA2010-21322-C03-02 and AIB-2010-DE-00227,
and is based on observations made with the William Herschel 
Telescope, operated on the island of La
Palma by the Isaac Newton Group in the Spanish Observatorio del Roque de los Muchachos
of the Instituto de Astrof\'isica de Canarias.
The work has also made use of the NASA/IPAC Extragalactic Database (NED) which is operated by 
the Jet Propulsion Laboratory, California Institute of Technology, under contract with the 
National Aeronautics and Space Administration, and of the SDSS database.
Funding for the SDSS and SDSS-II has been provided by the Alfred P. Sloan Foundation, 
the Participating Institutions, the National Science Foundation, the U.S. Department of Energy, 
the National Aeronautics and Space Administration, the Japanese Monbukagakusho, 
the Max Planck Society, and the Higher Education Funding Council for England. 

\bibliographystyle{mn2e} 
\bibliography{reference} 

\appendix

\section{All the maps galaxy by galaxy}\label{sec:app}
In this Section we put together the maps resulting from the analysis of the \sauron\ data for each galaxy, including
the stellar intensity and kinematics, gas intensity and kinematics, the measured line-strength indices,
and the mean luminosity-weighted ages and metallicities.

\begin{figure*}
\begin{center}
  \includegraphics[angle=0,width=.95\textwidth]{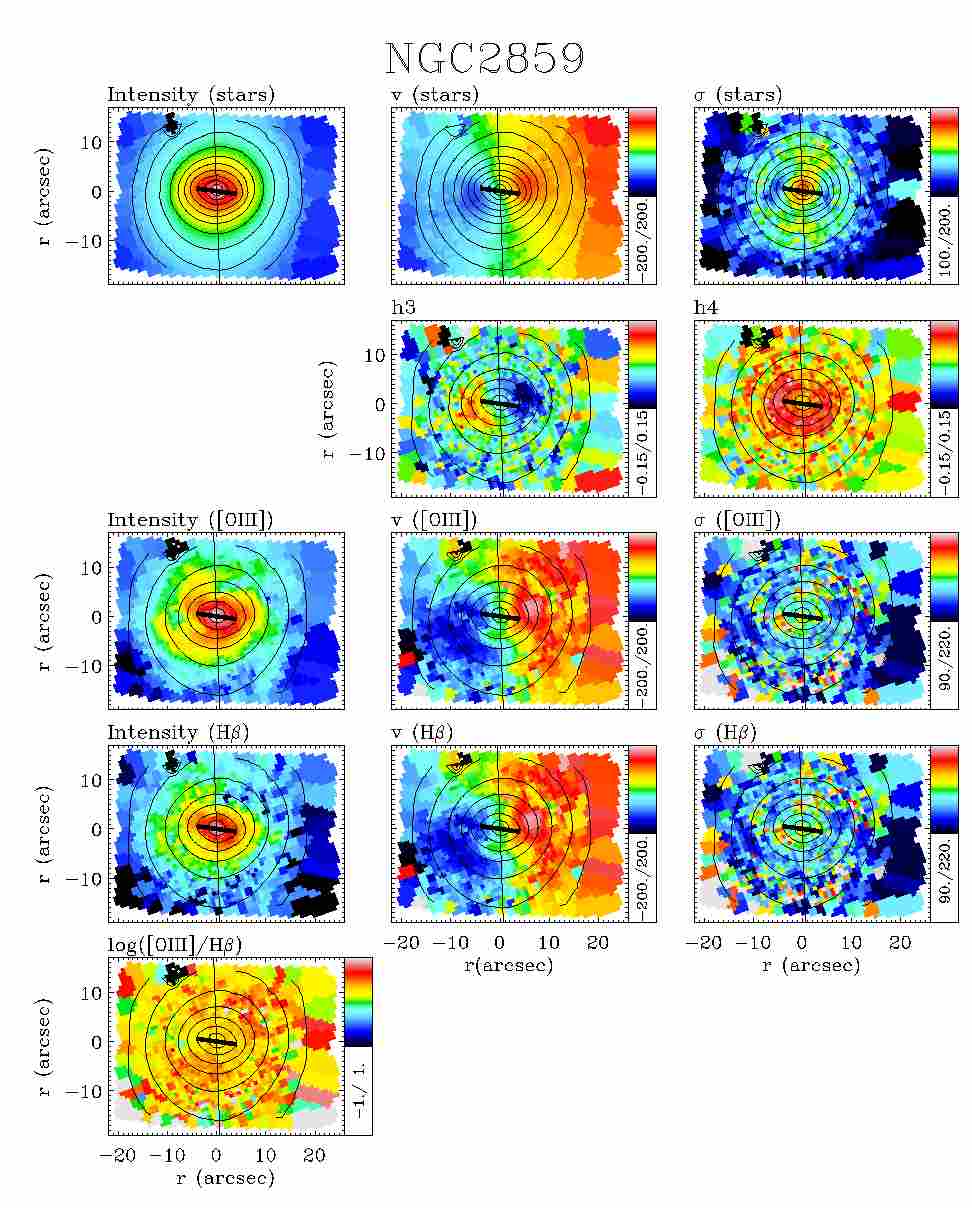}
  \caption{Kinematical analysis of the double-barred galaxy NGC\,2859.
The two top rows contain the stellar kinematics; from left to right and from top to bottom: 
stellar intensity (in arbitrary units), LOS velocity (km\,s$^{-1}$), velocity dispersion (km\,s$^{-1}$), 
h$_3$, and h$_4$ maps. The third and fourth rows 
represent the gas kinematics derived from the [OIII]
and H$\beta$ emission lines, respectively; from left to right: line intensity (arbitrary units),
velocity (km\,s$^{-1}$), and velocity dispersion (km\,s$^{-1}$). The bottom panel is the [OIII]/H$\beta$ gas
intensity ratio in logarithmic scale. 
For all the maps, we have overplotted the position angle and length of the inner bar (thick line), 
the position angle of the outer bar (thin line), 
and the contours of the reconstructed total intensity map.
The scale is 1\,arcsec\,$\sim$\,120\,pc.}
  \label{fig:n28591}
\end{center}
\end{figure*}

\begin{figure*}
\begin{center}
  \includegraphics[angle=0,width=.95\textwidth]{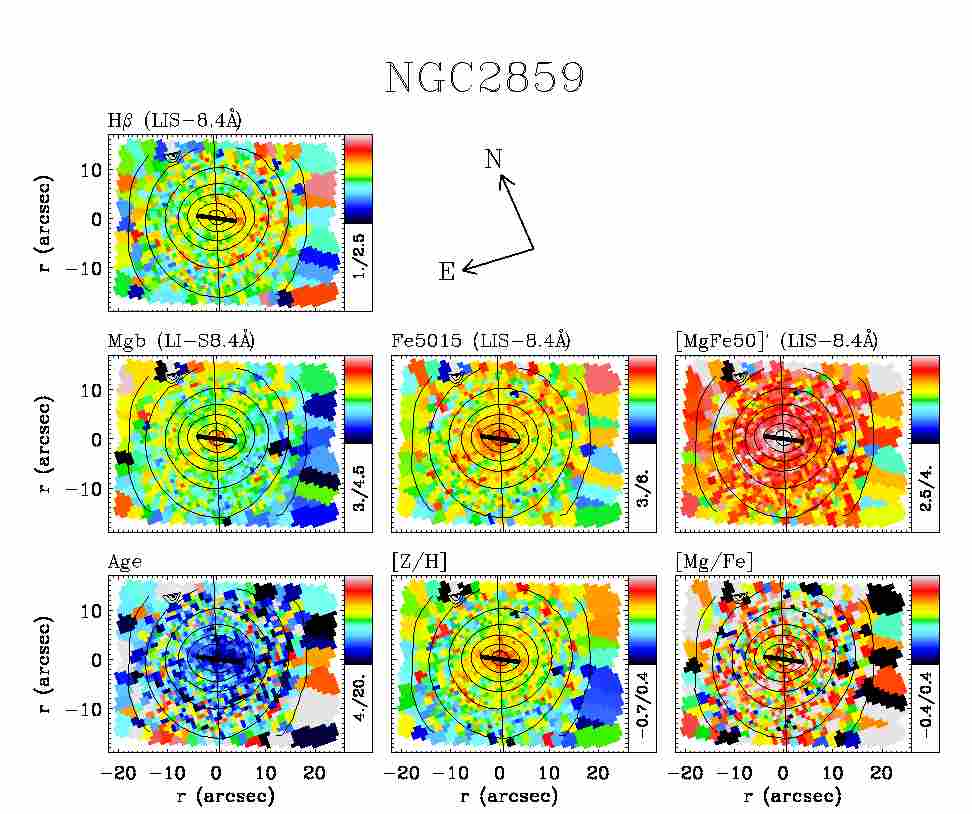}
  \caption{Stellar population analysis of the double-barred galaxy 
NGC\,2859. The first row
includes the map of the age-sensitive index H$\beta$,
whereas the middle row shows the maps corresponding to the metallicity indicators
Mg\emph{b}, Fe5015, and [MgFe50]$'$ (from left to right, respectively). 
All the indices are measured at a resolution of 8.4\AA\ (FWHM),
following the corresponding LIS system.
The bottom row contains the age (in Gyr), metallicity, and
$\alpha$-enhancement distributions, from left to right, respectively.
For all the maps, we have overplotted the position angle and length of the inner bar (thick line), 
the position angle of the outer bar (thin line), and the contours of the reconstructed total intensity map. 
The relative orientation on the sky is also indicated.
The scale is 1\,arcsec\,$\sim$\,120\,pc.}
  \label{fig:n28592}
\end{center}
\end{figure*}

\begin{figure*}
\begin{center}
  \includegraphics[angle=0,width=.95\textwidth]{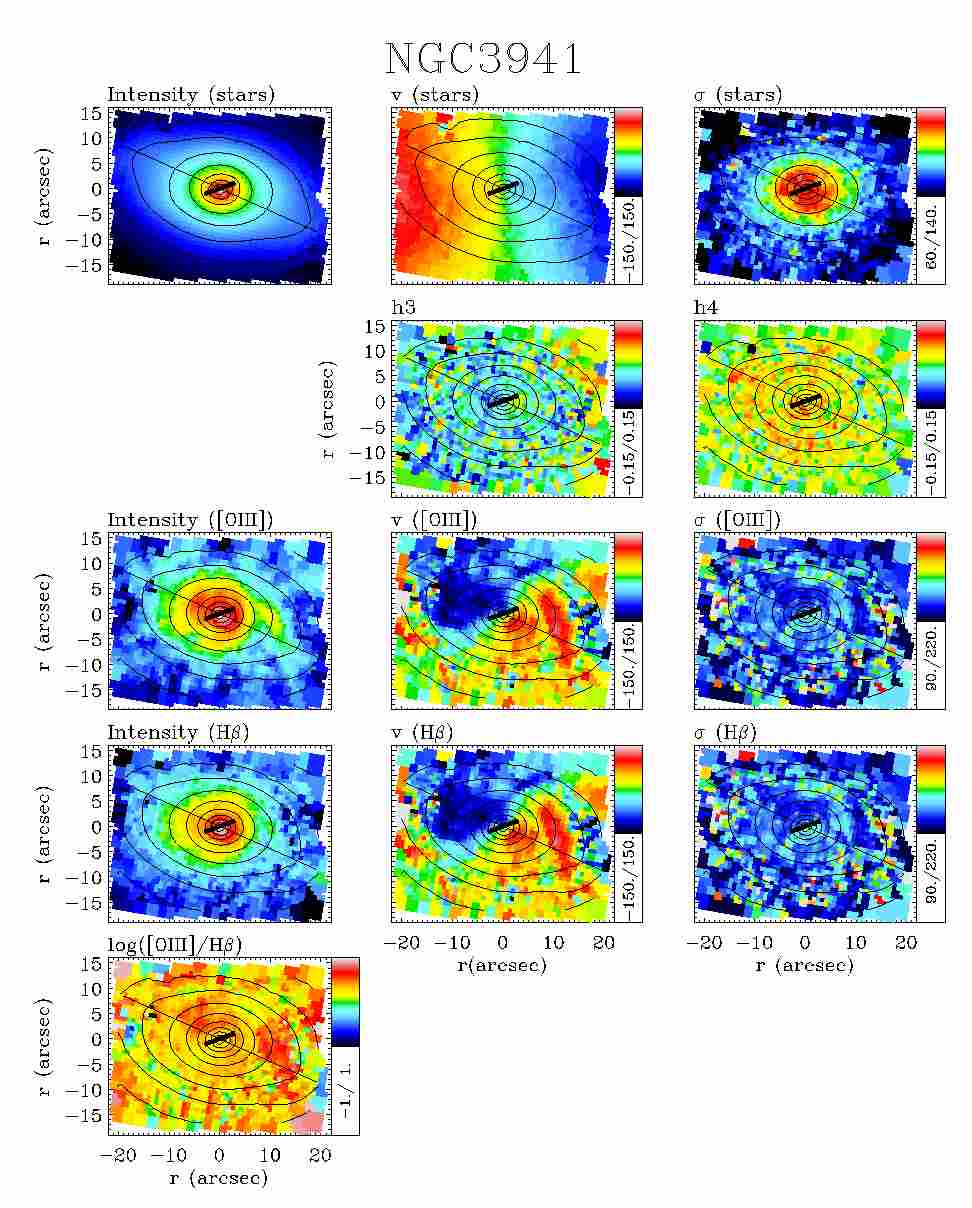}
  \caption{Same as Figure \ref{fig:n28591} but for NGC\,3941.
The scale is 1\,arcsec\,$\sim$\,90\,pc.}
  \label{fig:n39411}
\end{center}
\end{figure*}

\begin{figure*}
\begin{center}
  \includegraphics[angle=0,width=.95\textwidth]{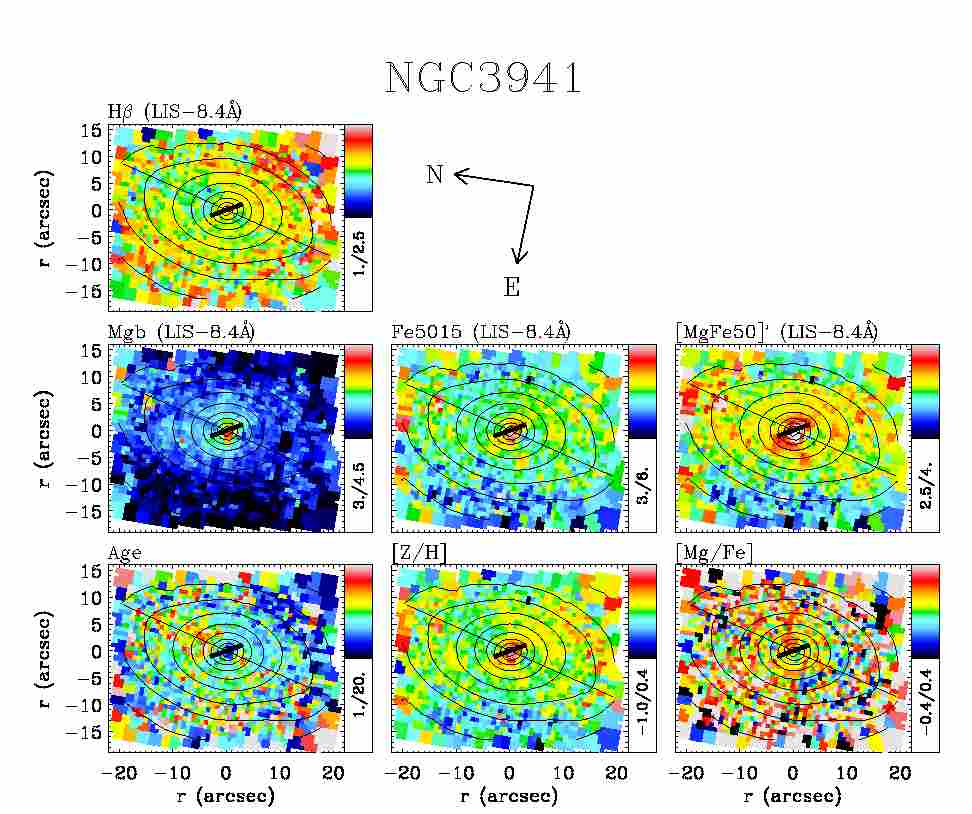}
  \caption{Same as Figure \ref{fig:n28592} but for NGC\,3941.
The scale is 1\,arcsec\,$\sim$\,90\,pc.}
  \label{fig:n39412}
\end{center}
\end{figure*}

\begin{figure*}
\begin{center}
  \includegraphics[angle=0,width=.95\textwidth]{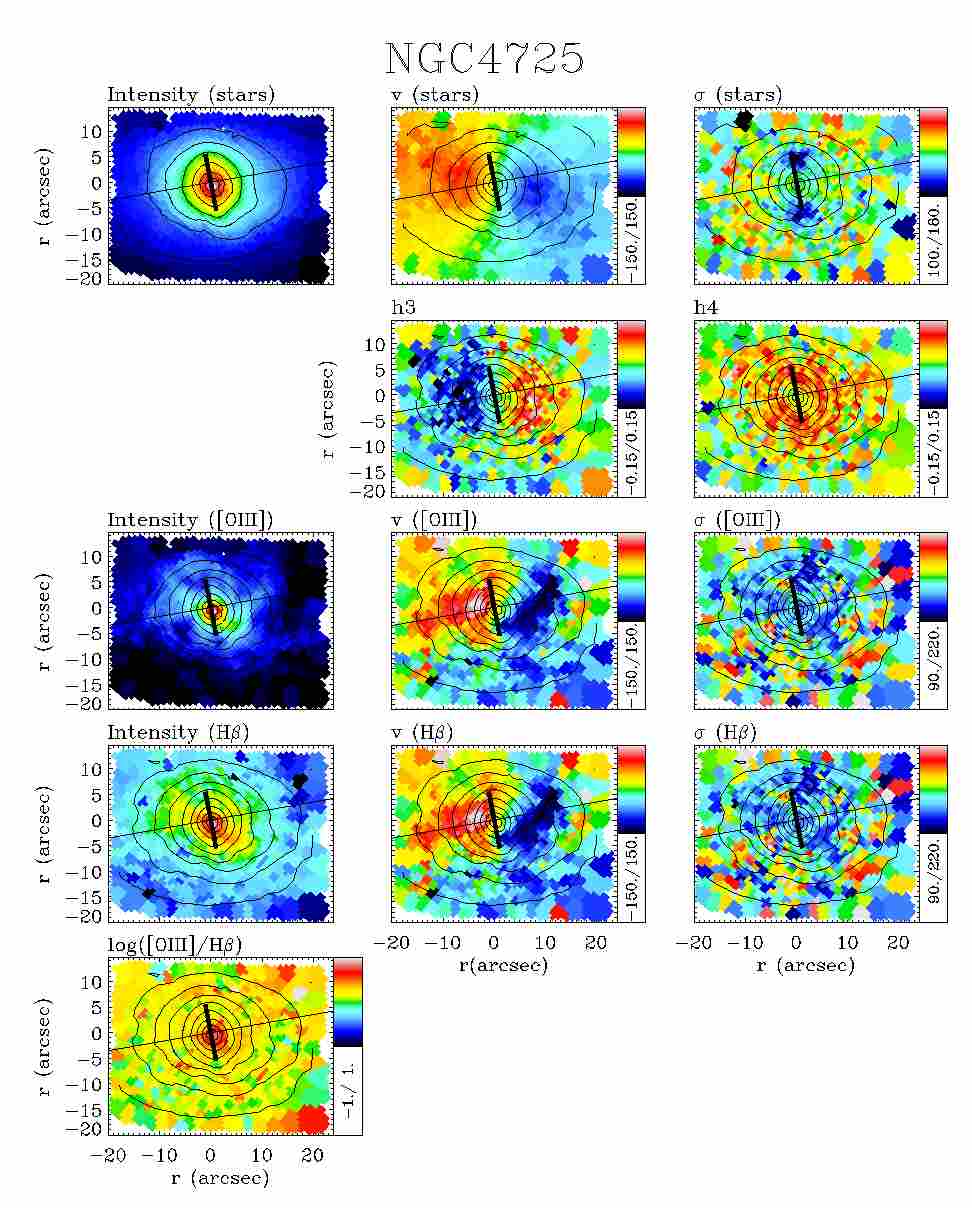}
  \caption{Same as Figure \ref{fig:n28591} but for NGC\,4725.
The scale is 1\,arcsec\,$\sim$\,60\,pc.}
  \label{fig:n47251}
\end{center}
\end{figure*}

\begin{figure*}
\begin{center}
  \includegraphics[angle=0,width=.95\textwidth]{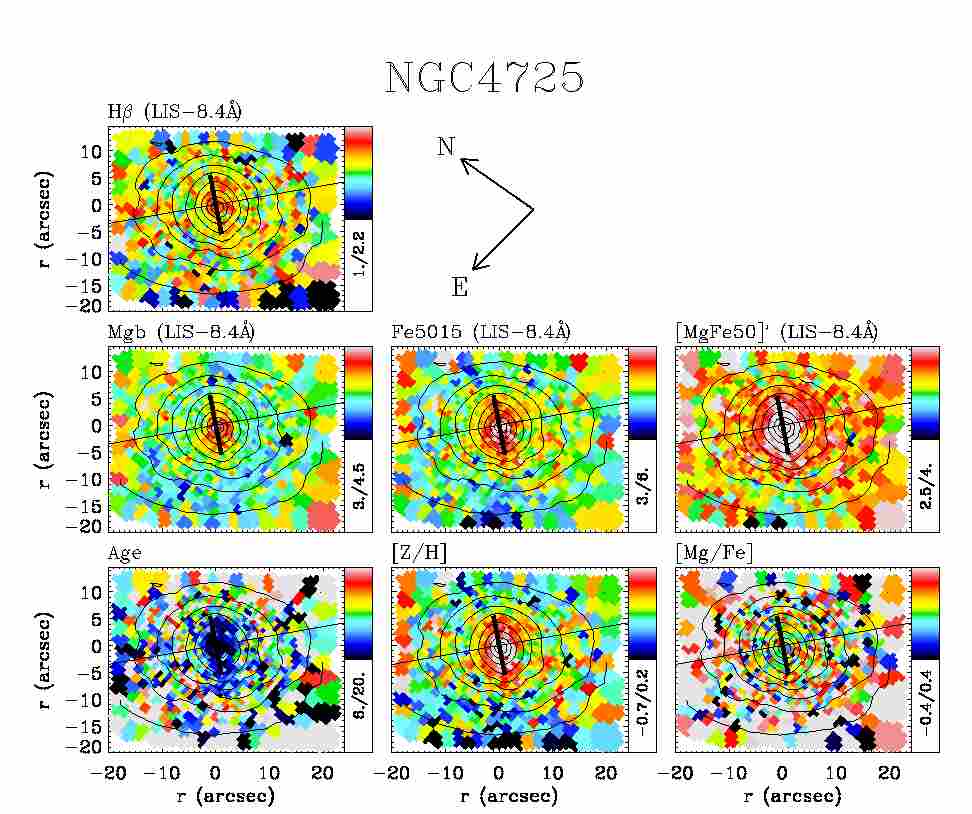}
  \caption{Same as Figure \ref{fig:n28592} but for NGC\,4725.
The scale is 1\,arcsec\,$\sim$\,60\,pc.}
  \label{fig:n47252}
\end{center}
\end{figure*}

\begin{figure*}
\begin{center}
  \includegraphics[angle=0,width=.95\textwidth]{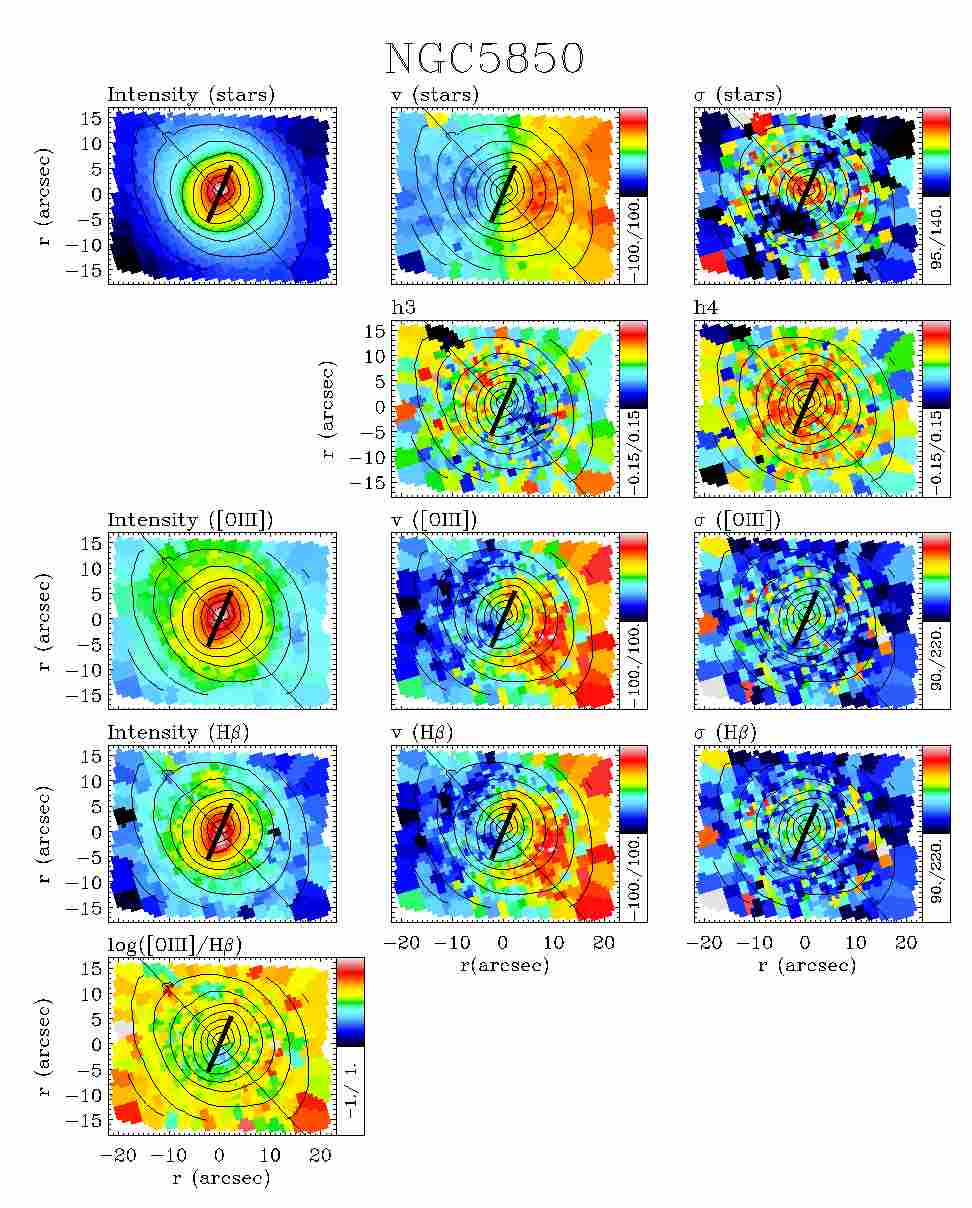}
  \caption{Same as Figure \ref{fig:n28591} but for NGC\,5850.
The scale is 1\,arcsec\,$\sim$\,140\,pc.}
  \label{fig:n58501}
\end{center}
\end{figure*}

\begin{figure*}
\begin{center}
  \includegraphics[angle=0,width=.95\textwidth]{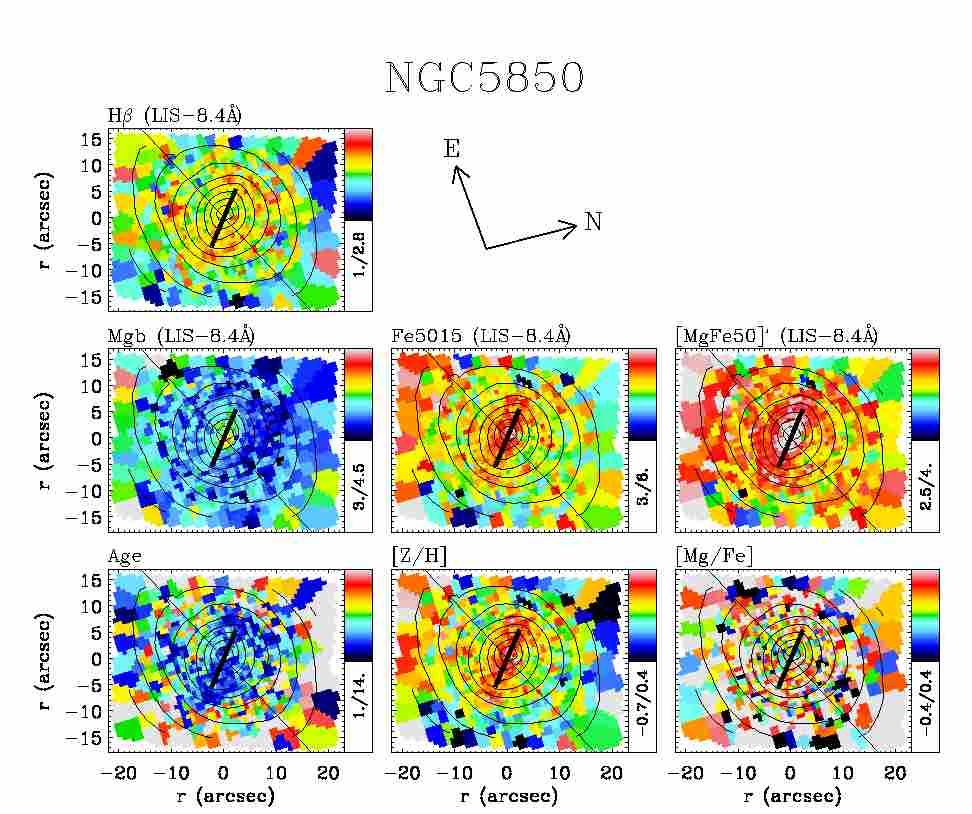}
  \caption{Same as Figure \ref{fig:n28592} but for NGC\,5850.
The scale is 1\,arcsec\,$\sim$\,140\,pc.}
  \label{fig:n58502}
\end{center}
\end{figure*}

\section{Index--index diagrams}\label{sec:appgrids}
The following Figures contain the results from the index--index diagram analysis:
the model predictions for an age-sensitive index are plotted versus the predictions
for a metallicity-sensitive index, and the same measurements for the data are 
overplotted in order to study the age and metallicity of our galaxies.

\begin{figure*}
\begin{center}
\includegraphics[width=\linewidth]{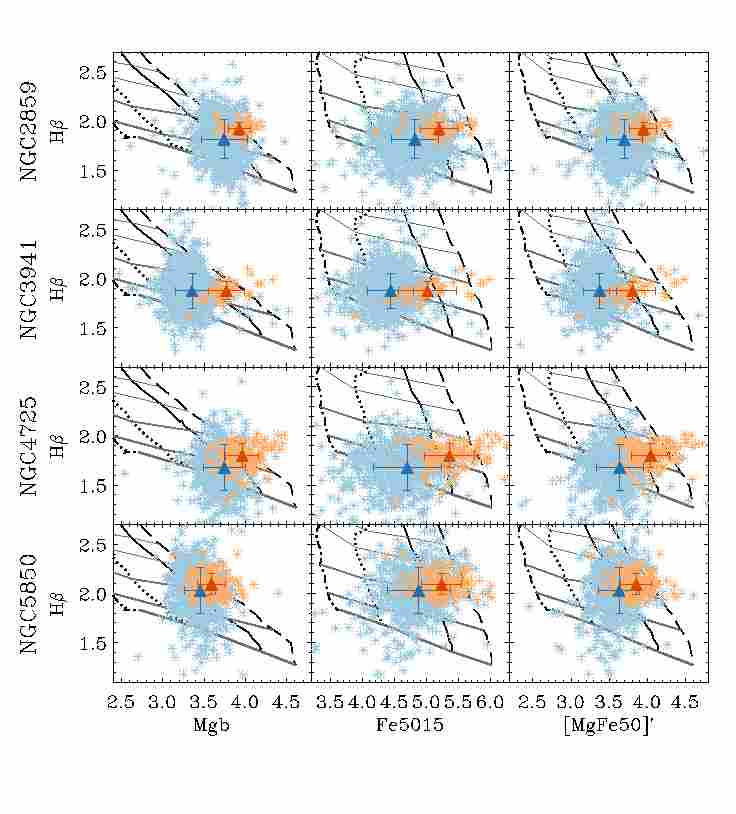}
\caption{Age indicator H$\beta$ versus the metallicity
indicators Mg\emph{b}, Fe5015, and [MgFe50]$'$ (from left to right, respectively) for the four
double-barred galaxies: NGC\,2850, NGC\,3941, NGC\,4725, and NGC\,5850 (from top to bottom, respectively). 
The grids correspond to the SSP
models by \citet{Vazdekisetal2010}. The solid lines 
represent different ages increasing from top to bottom (2.5, 3, 5.6, 10, and 18 Gyr, respectively),
whereas the almost vertical lines indicate different metallicities increasing from left to right
([Z/H]=\,-0.7, -0.4, 0.0, and 0.2, respectively). The asterisks are
the measurements for all the Voronoi bins of each galaxy, distinguishing
the bins corresponding to the inner bar (orange asterisks) from those
corresponding to the outer bar (blue asterisks). The triangles represent the
average measurements for the inner bar (red triangles), and the rest of the
galaxy (blue triangles). Both models and galaxy measurements 
are carried out in the LIS-8.4\,\AA\ system.}
  \label{fig:grid2859}
\end{center}
\end{figure*}

\label{lastpage}

\end{document}